\documentclass[aps,prd,superscriptaddress,nofootinbib,amsmath,amsfonts,preprintnumbers,notitlepage,10pt,english]{revtex4-1}
\usepackage{amsmath}
\usepackage{amssymb}
\usepackage{babel}

\makeatletter

\@ifundefined{textcolor}{}
{%
 \definecolor{BLACK}{gray}{0}
 \definecolor{WHITE}{gray}{1}
 \definecolor{RED}{rgb}{1,0,0}
 \definecolor{GREEN}{rgb}{0,1,0}
 \definecolor{BLUE}{rgb}{0,0,1}
 \definecolor{CYAN}{cmyk}{1,0,0,0}
 \definecolor{MAGENTA}{cmyk}{0,1,0,0}
 \definecolor{YELLOW}{cmyk}{0,0,1,0}
 }

\@ifundefined{textcolor}{}{%
 \definecolor{BLACK}{gray}{0}
 \definecolor{WHITE}{gray}{1}
 \definecolor{RED}{rgb}{1,0,0}
 \definecolor{GREEN}{rgb}{0,1,0}
 \definecolor{BLUE}{rgb}{0,0,1}
 \definecolor{CYAN}{cmyk}{1,0,0,0}
 \definecolor{MAGENTA}{cmyk}{0,1,0,0}
 \definecolor{YELLOW}{cmyk}{0,0,1,0}
 }
\usepackage{array,multirow,graphicx}
\usepackage{dcolumn}
\usepackage{txfonts}
\usepackage{newlfont}
\usepackage{times}
\usepackage{bm}
\usepackage[colorlinks,citecolor=blue,urlcolor=blue,linkcolor=blue]{hyperref}
\usepackage[figtopcap]{subfigure}
\usepackage{color}

\makeatother

\begin{document}

\date{\today}

\title{Charged spherically symmetric black holes in $f(R)$ gravity and their stability analysis}
\author{  G.  G. L. Nashed$^{1,2}$ and S. Capozziello$^{3,4,5,6}$ }
\affiliation{Centre for Theoretical Physics, The British University in Egypt, P.O. Box 43,\\ El Sherouk City, Cairo 11837, Egypt\\
$^2$Department of Mathematics, Faculty of Science, Ain Shams University, Cairo 11566, Egypt\\
$^3$Dipartimento di Fisica ``E. Pancini``, Universit\'a di Napoli ``Federico II'',
Complesso Universitario di Monte Sant' Angelo, Edificio G, Via Cinthia, I-80126, Napoli, Italy\\
$^4$ Istituto Nazionale di Fisica Nucleare (INFN),  Sezione di Napoli,
Complesso Universitario di Monte Sant'Angelo, Edificio G, Via Cinthia, I-80126, Napoli, Italy\\
$^5$Gran Sasso Science Institute, Viale F. Crispi, 7, I-67100, L'Aquila, Italy.\\
$^6$Laboratory for Theoretical Cosmology, \\Tomsk State University of Control Systems and Radioelectronics (TUSUR), 634050 Tomsk, Russia.
\\
}
%\begin{center}
%{\bf  Charged spherically symmetric black hole solutions in $f(R)$ Gravity}
%\end{center}
%\begin{center}
%{\bf Gamal G.L. Nashed}\footnote{ Corresponding author, E-mail: nashed@bue.edu.eg}
%\end{center}

%\bigskip

%\centerline{\it Centre  for Theoretical Physics, The British
%University in  Egypt} \centerline{\it Sherouk City 11837, P.O. Box
%43, Egypt\footnote{ Mathematics Department, Faculty of Science, Ain
%Shams University, Cairo, 11566, Egypt\\
%\hspace*{.2cm} Egyptian Relativity Group (ERG) URL:
%http://www.erg.eg.net}}
\begin{abstract}
A new class of analytic charged spherically symmetric black hole solutions, which behave asymptotically as flat or (A)dS spacetimes, is derived for  specific classes of $f(R)$  gravity, i.e.,  $f(R)=R-2\alpha\sqrt{R}$ and $f(R)=R-2\alpha\sqrt{R-8\Lambda}$, where $\Lambda$ is the cosmological constant. These black holes are characterized by the dimensional parameter  $\alpha$ that makes solutions deviate from the standard solutions of general relativity. The Kretschmann scalar and  squared Ricci tensor  are shown to depend on the parameter $\alpha$ which is not allowed to be  zero.  Thermodynamical quantities, like entropy, Hawking temperature,  quasi-local energy and the Gibbs free energy are calculated.   From these calculations, it is possible to   put a constrain on the dimensional parameter $\alpha$ to have $0<\alpha<0.5$, so that all  thermodynamical quantities have a physical meaning.  The interesting result of these calculations is the possibility of a negative black hole entropy.  Furthermore, present calculations show that for negative energy, particles inside a black hole, behave as if they have a negative entropy. This fact gives rise to  instability for  $f_{RR}<0$.  Finally, we study the linear metric perturbations of the derived black hole solution. We show that for the odd-type modes, our black hole is always stable and has a radial speed with fixed value equal to $1$. We also, use the geodesic deviation to derive further stability  conditions.

\end{abstract}


%\begin{center}
\maketitle
\section{\bf Introduction}
%\end{center}

 Challenging problems ranging from  quantum gravity to dark energy (DE) and dark matter (DM)   give support  to search for other gravitational theories beyond the standard  Einstein general relativity (GR). Actually,  GR has many unsolved issues like singularities, the nature of DE and DM, etc.  All these issues encourage scientists to modify GR or extend it in view of addressing shortcomings at UV and IR scales \cite{Capozziello:2011et}. In other words, viable modified/extended theories should be compatible with the current experimental constraints and should give motivations on  issues in quantum gravity and cosmology. Thus, it is straightforward to extend directly GR considering it as a limit of a more general theory of gravitation\footnote{It is worth saying that {\it modified gravity} means that GR is not recovered but equivalent scheme as the Teleparallel Equivalent General Relativity (TEGR) can be recovered. Extended gravity means that in a given limit, or for a given choice  GR is recovered. For a discussion see \cite{Cai:2015emx}.}. Among the possible extensions of GR,  the so-called $f(R)$ gravity  generalizes the Einstein-Hilbert action by substituting the Ricci scalar $R$ by an analytic differentiable function. The fundamental reasons for this approach come out from the formulation of any quantum field theory on curved spacetime \cite{Birrell:1982ix}. $f(R)$ gravity have some important applications, like the  Starobinsky model,  $f(R)=R+\alpha R^2, \alpha>0$ , which is  successful  to explain  inflationary behavior of early universe \cite{1980PhLB...91...99S,Nojiri:2003ft,Nojiri:2003ni}. Furthermore $f(R)$ gravity  is capable of explaining the observed cosmic  acceleration  without assuming the cosmological constant.  Possible toy models have the form              ${\displaystyle f(R)=R-\frac{\beta}{R^n}}$ where $\beta$ and $n$ have positive values \cite{Capozziello:2003gx,Capozziello:2002rd,Carroll:2003wy}. Nevertheless, this model suffers from instability problems because of the second derivative of the function $f$ that has a negative value, i.e., $f_{RR}<0$ \cite{Dolgov:2003px,Faraoni:2006sy,Sawicki:2007tf,Song:2006ej,Chiba:2003ir,PhysRevLett.95.261102,PhysRevD.72.083505,PhysRevD.74.121501,Jana:2018djs}. Later this problem has been tackled  \cite{Hu:2007nk}  and cosmological stable models have been achieved  using some limitations on the parameter space. There are many viable cosmological models constructed using $f(R)$ \cite{Starobinsky:2007hu,Nojiri:2006gh}. Finally $f(R)$  models  give interesting results for structure formation, as the modification of the spectra of galaxy clustering, CMB, weak lensing, etc. \cite{PhysRevD.77.046009,PhysRevD.77.023503,PhysRevD.73.123504,PhysRevD.75.084010,PhysRevD.76.063517,PhysRevD.78.043002,Nojiri:2007cq,Nojiri:2007as,Capozziello:2018ddp}.  There are many  applications of $f(R)$ from the astrophysical point of view \cite{PhysRevLett.101.061103,PhysRevD.80.064002,PhysRevD.82.064033,PhysRevD.92.064015,Bamba:2008ut,Capozziello:2006dj,PhysRevD.93.023501,Capozziello:2012ie}, for general reviews of $f(R)$ gravity,  see \cite{Capozziello:2011et, Nojiri:2010wj,Nojiri:2006ri,Nojiri:2017ncd}.

From the viewpoint of mathematics, modified/extended gravity poses the issue to establish or modified  well-known facts of GR like the stability of solutions, initial value problem and the problem of deriving new black hole solutions \cite{Capozziello:2009up,Faraoni_2009,Olmo:2008nf,PhysRevD.78.064017,Capozziello:2009pi}. As it is well-known, in addition to the cosmological solutions, there exist axially symmetric as well as spherical ones that could have a main role in several astrophysical problems spanning from black hole solutions to galactic nuclei. Modified gravitational theories must include black hole solutions like Schwarzschild-like in order to be compatible with GR results and, in principle, must give new black hole solutions that might have physical interest. According to this fact, the way to find out exact or approximate black hole solutions is highly important to investigate if observations can be matched  to modified/extended  gravity \cite{Capozziello:2007wc,Capozziello:2009jg}.

In the framework of $f(R)$ gravity there is a specific interest for  spherically symmetric black hole solutions.  They have been derived using constant Ricci scalar \cite{PhysRevD.74.064022}. Moreover, spherically symmetric black hole solutions, including perfect fluid matter, have been  analyzed \cite{PhysRevD.76.064021}. Additionally, by using the method of Noether symmetry, many spherically symmetric black holes have been derived \cite{PhysRevD.76.064021}. Hollenstein and Lobo \cite{Hollenstein:2008hp} derived exact solutions of static spherically symmetric spacetimes in $f(R)$  coupled to  non-linear electrodynamics. For the readers interested in the static black holes we refer to \cite{PhysRevD.90.084011,Hendi:2011eg,Nashed:2005kn,PhysRevD.85.044012,Nojiri:2014jqa,Nojiri:2017kex,PhysRevD.84.084006,Awad:2017tyz,Cembranos:2011sr,Hendi:2011eg,Nashed:2006yw,PhysRevD.80.104012,Azadi:2008qu,Nashed:2009hn,Nashed:2008ys,Capozziello:2007id,PhysRevD.76.024020,Nashed:2007cu,Nashed:2015pda,PhysRevD.91.104004,2013CQGra..30l5003N,Nashed:2016tbj,2014CaJPh..92...76H,Hendi:2014mba,Nashed:uja,Nashed:2015qza,PhysRevD.81.124051,Nashed:2018piz,2018EPJP..133...18N,2018IJMPD..2750074N,PhysRevD.80.121501,Hendi:2012nj,Myrzakulov:2015kda,PhysRevD.92.124019,Hanafy:2015yya,PhysRevLett.114.171601,PhysRevD.75.027502} and the references therein.  Using the Lagrangian multiplier, new analytic solutions with dynamical Ricci scalar have been derived \cite{Sebastiani:2010kv}. It is the purpose of the present study, by using the field equation of $f(R)$, to generalize these black hole solutions \cite{Sebastiani:2010kv} and derive new charged black hole solutions with dynamical Ricci scalar asymptotically converging towards flat or (A)dS spacetimes.

 Gravitational stability of a black hole solution is considered as  a main problem for checking the adequateness of any
black hole solutions \cite{Konoplya:2011qq,PhysRevD.1.2870}. However, the stability analysis appears not directly to be applicable to $f(R)$ black
hole solutions because   it  involves fourth-order derivative
terms in the linearized equations \cite{Psaltis:2007cw,Barausse:2008xv}.  In that case, it is necessary that the black holes  are free
from tachyon and ghost  instabilities  that would come into the game as soon as one is considering $f(R)$ gravity \cite{DeFelice:2011ka}. Therefore, one may transform $f(R)$ gravity into the corresponding scalar-tensor theory to remove the
fourth-order derivative terms  \cite{Olmo:2005zr}. It was suggested that the stability of black hole solutions does not rely on the frame due to the fact that  it is a classical
solution that is considered as the ground state \cite{PhysRevD.68.024028}.   It is well known that a non-minimally coupled scalar makes the linearized
GR field equations around the black hole very intricate when   compared to a minimally
coupled scalar in the context of GR \cite{PhysRevD.34.333}. Due to this intricacy, some people have
used conformal transformations to find the corresponding theory in the Einstein frame
where a minimally coupled scalar appears. Taking into account these difficulties, several  perturbation studies on the black holes in different  modified gravitational theories have been developed. See, for example, \cite{DeFelice:2011ka,Moon:2011sz,Capozziello:2007ms,Moon:2011fw,Myung:2014nua}.

The paper is organized  as follows. In Sec. \ref{S2}, a summary of   Maxwell-$f(R)$ gravity is  provided. In Sec. \ref{S3}, a spherically symmetric ansatz is applied to the field equation of Maxwell-$f(R)$ theory and an exact solution is derived. In Sec. \ref{S4}, the same spherically symmetric ansatz is applied to the field equation of Maxwell-$f(R)$ theory that includes a cosmological constant. Solving the resulting differential equations,  we derive a  new black hole solution that behaves as (A)dS. In Sec. \ref{S55}, the characteristic properties of these black holes are analyzed. In Sec. \ref{S66},   thermodynamical quantities like entropy, quasi-local energy, Hawking temperature, Gibbs energy are calculated. We  show that the entropy of the derived black hole solutions are not proportional to the horizon area and show some regions of the parameter space where the entropy becomes negative. The main reason for this result is due to  the parameter $\alpha$  related to the higher order correction.   In Sec.  \ref{S616} we study the linear stability using the odd perturbations to the black holes derived in Secs. \ref{S3}  and \ref{S4}. Furthermore,  in  Sec. \ref{S616},  we derive the stability conditions considering the  geodesic
motion. In Sec. \ref{S77}, we discuss the main results of the present study and draw conclusions.

\section{  Maxwell--$f(R)$ gravity}\label{S2}
The   theory of gravity that will be considered  in this work is the $f(R)$ gravity which was first taken into account   in \cite{1970MNRAS.150....1B}. See also \cite{Capozziello:2011et,2010deto.book.....A,Capozziello:2003gx,Capozziello:2002rd,Carroll:2003wy}:
\begin{eqnarray} \label{a1}
{\mathop{\mathcal{ S}}}:={\mathop{\mathcal{ S}}}_g+{\mathop{\mathcal{ S}}}_{_{E.M.}},\end{eqnarray}
where ${\mathop{\mathcal{ S}}}_g$ is the gravitational action given by:
\begin{eqnarray} \label{a2} {\mathop{\mathcal{ S}}}_g:=\frac{1}{2\kappa} \int d^4x \sqrt{-g} (f(R)-\Lambda),\end{eqnarray}
where $\Lambda$ represents the cosmological constant,  $R$ is the Ricci scalar, $\kappa$  is the gravitational constant, $g$ is the determinant of the metric and $f(R)$ is an analytic  differentiable  function. In this study
${\mathop{\mathcal{S}}}_{E.M.} $ is  the action  of the non-linear electrodynamics field which takes the form:
 \begin{eqnarray}\label{a3} {\mathop{\mathcal{S}}}_{
E.M.}:=-\frac{1}{2}F^{2s},  \end{eqnarray} where  $s\geq 1$ is an arbitrary parameter that is  equal to one for the standard Maxwell theory and $F^2=F_{\mu \nu}F^{\mu\nu}$, where $F_{\mu \nu} =2A_{[\mu, \nu]}$  with $A_\mu$ being  the gauge potential 1-form and the comma denotes the ordinary differentiation\footnote{The square brackets represent the  anti-symmetrization, i.e. $A_{[\mu, \nu]}=\frac{1}{2}(A_{\mu, \nu}-A_{\nu ,\mu})$ and the symmetric one is represented by $A_{(\mu, \nu)}=\frac{1}{2}(A_{\mu, \nu}+A_{\nu ,\mu})$.} \cite{Hendi:2012nj}.

 The  field equations of $f(R)$ gravitational theory can be obtained by carrying out the variations of the action given by Eq. (\ref{a1}) with respect to the metric tensor $g_{\mu \nu}$ and the strength tensor $F$ that
yield the following form of the  field equations \cite{2005JCAP...02..010C,Koivisto:2005yc}:
%\newpage
\begin{eqnarray} \label{f1}
I_{\mu \nu}=R_{\mu \nu} f_R-\frac{1}{2}g_{\mu \nu}f(R)-2g_{\mu \nu}\Lambda +g_{\mu \nu} \Box f_R-\nabla_\mu \nabla_\nu f_R-8\pi T_{\mu \nu}\equiv0,\end{eqnarray}
\begin{equation}\label{f2}
\partial_\nu \left( \sqrt{-g} {\textrm F}^{\mu \nu} F^{s-1} \right)=0, \end{equation}
with $R_{\mu \nu}$ being the Ricci tensor defined by
 \[R_{\mu \nu}=R^{\rho}{}_{\mu \rho \nu}=  2\Gamma^\rho{}_{\mu [\nu,\rho]}+2\Gamma^\rho{}_{\beta [\rho}\Gamma^\beta{}_{\nu] \mu},\]
 where $\Gamma^\rho{}_{\mu \nu}$ is the Christoffel symbols of second kind.
 The d'Alembert operator $\Box$ is defined as $\Box= \nabla_\alpha\nabla^\alpha $ where $\nabla_\alpha V^\beta$ is the covariant derivatives of the vector $V^\beta$ and ${\displaystyle f_R=\frac{df(R)}{dR}}$.   In this study $T_{\mu \nu}$ is defined as
 \begin{eqnarray} T_{\mu \nu}:=\frac{1}{4\pi}\left(s{ \textrm g}_{\rho
\sigma}{{\textrm  F}_\nu{}^\rho}{{{\textrm F}}_\mu}^{\sigma}F^{s-1}-\displaystyle{1 \over 4}  {\textrm g}_{\mu \nu} F^{2s}\right), \end{eqnarray} which is the energy momentum-tensor of the non-linear electrodynamic field. When $s=1$, we get the standard energy-momentum tensor of Maxwell field.

The trace of  equation  (\ref{f1}) is:
\begin{eqnarray} \label{f3}
Rf_R-2f(R)-8\Lambda+3\Box f_R=T \qquad \textrm{where} \qquad T=F^s(sF-F^s)\,,\end{eqnarray}
It is worth noticing that, for $s=1$, it is $T=0$. This property means that the Maxwell field is conformally invariant.
In the following, we are going to assume some form of the field Eqs. (\ref{f1}) without and with a cosmological constant to derive exact solutions that asymptotically behave as flat or (A)dS spacetimes.

%%%%%%%%%%%%%%%%%%%%%%%%%%%%%%%%%%% Section 3 %%%%%%%%%%%%%%%%%%%%%%%%%%%%%%%%%%%%%%%%
\section{An exact charged black hole solution  }\label{S3}
%%%%%%%%%%%%%%%%%%%%%%%%%%%%%%%%%%%%%%%%%%%%%%%%%%%%%%%%%%%%%%%%%%%%%%%%%%%%%%%%%%%%%%
Let us derive a charged black hole solution adopting the model  $f(R)=R-2\alpha\sqrt{R}$. To this aim, we are going to use the following spherically symmetric ansatz\footnote{The reason to take the ansatz (\ref{met}) is to be able to find an exact solution. Other forms make the field equations very complicated and not easy to solve.}:
\begin{eqnarray} \label{met}
& &  ds^2=B(r)dt^2-\frac{dr^2}{B(r)}-r^2d\Omega^2,  \end{eqnarray}
where $d\Omega^2=d\theta^2+\sin^2\theta$ is the line element on the unit sphere. The Ricci scalar of the metric (\ref{met}) has the form
\begin{eqnarray}\label{r1}
& & R=\frac{2-r^2B''-4rB'-2B}{r^2}.
\end{eqnarray}
Applying the  ansatz (\ref{met}) to Eqs. (\ref{f1}), (\ref{f2}) and (\ref{f3}), after using (\ref{r1}) and  putting the parameter  $s=1$, we get the following non-vanishing field equations\footnote{Here and through all this study  $B\equiv B(r)$, $B'=\frac{dB(r)}{dr}$, $B''=\frac{d^2B(r)}{dr^2}$, etc. Also in this application we  put $\Lambda=0$ }:
\newpage
\begin{eqnarray} \label{df1}
& &
I_t{}^t=\frac{1}{2r^{10}\sqrt{R^9}}\Big\{r^6\sqrt{R^7}[r^4BB''''+B'''(1/2r^4B'+6r^3B)+2r^2B''(B+rB')-r^2B'^2+2B'(r-3rB)+4B(B-1)]+r^4\sqrt{R^5}
\Big[4r^2B'^2\nonumber\\
& &+r^6RBB''''-r^6BB'''^2+1/2r^3B'''\{r^2B''(rB'-4B)
+2B'(29rB-r)+40B(B-1)\}+2r^4B''^2(r^2q''-6B+2rB'-1)+r^2B''\nonumber\\
& &\times[23r^2B'^2+2rB'(23B+8r^2q'^2-13)+8(B-1)(6B+r^2q'^2-1)]+28r^3B'^3+
B'^2(34r^2B+32r^4q'^2-54r^2)+4rB'(B-1)\nonumber\\
& &\times(7B+r^2q'^2-9)+8(B-1)^2[r^2q'^2-1]\Big]+r^4R^2\Big[B\sqrt{R}[4r^2B''+r^3B'''-2rB'+4(1-B)]^2+\alpha\Big\{r^6RBB''''-
3/2r^6BB'''^2\nonumber\\
& &+1/2r^3B'''[r^2B''(rB'-12B)+4r^2B'^2+2B'(31rB-r)+48B(B-1)]-r^6B''^3-100r^3B'^3+2r^2B'^2[96B-85]+B''^2\Big(8r^4\nonumber\\
& &-12r^5B'-30r^4B\Big)+r^2B''[57r^2B'^2+14rB'[3B-5]+20+4B(4-3B)]
-4rB'(B-1)[27B-23]-16(B-1)^2(2B-1)\Big\}\Big]\Big\}=0,\nonumber\\
& & I_r{}^r=- \frac{1}{4r^8\sqrt{R^7}}\Big\{r^4\sqrt{R^5}\Big[(rB'+4B)[4(1-B)-2rB'+r^2(4B''+rB''')]-\Big(r^3B'''[rB'+4B]+4r^2B''[5B+r^2q'^2+2rB'-1]\nonumber\\
& &
+14r^2B'^2+B'[16r^3q'^2+12rB-20r]+8(B-1)[r^2q'^2-B-1]\Big)\Big]-\alpha r^4R^2\Big(r^3B'''[4B+rB']-2r^4B''^2+B''\Big[4r^2(B+3)\nonumber\\
& &-16r^3B'\Big]-50r^2B'^2+4rB'(15-17B)-16[2B-1][B-1]\Big)\Big\}=0,\nonumber\\
 & & I_\theta{}^\theta=I_\phi{}^\phi=
 \frac{-1}{2r^{10}\sqrt{R^9}}\Big\{r^6\sqrt{R^7}\Big[r^4BB''''+r^3B'''(rB'+5B)+2r^2B''(2rB'-B)+2rB'(2-3B)-2r^2B'^2+8B(B-1)\Big]\nonumber\\
& &-r^5\sqrt{R^5}\Big[r^5BRB''''-r^5BB'''^2
 +r^2B'''\Big\{r^2B''(rB'-3B)+4r^2B'^2+2rB'(13B-2)+18B(B-1)\Big\}+r^5B''^3-2r^3B''^2(r^2q'^2+2\nonumber\\
& &-7rB'+7B)+2rB''\Big(23r^2B'^2-rB'[14+8r^2q'^2-17B]+2(B-1)[1-10B+2r^2q'^2]\Big)
 +24r^2B'^3-4rB'^2(3+8r^2q'^2)\nonumber\\
& &+4B'(B-1)(3B-r^2q'^2)-8rq'^2(B-1)^2\Big]+r^4R^2\Big[\sqrt{R}(B[(B-1)-r^3B'''-4r^2B''+2rB']^2)+\alpha\Big(r^6BB''''R+3/2r^6BB'''^2\nonumber\\
& &
 -r^3B'''[r^2B''\{rB'-7B\}+4r^2B'^2+2rB'[14B-1]+22B(B-1)]+2r^6B''^3+2r^4B''^2[18B+9rB'-5]+2r^2B''[33r^2B'^2\nonumber\\
& &+rB'(27B-34)-2(9B^2-5B-4)]
 +104r^3B'^3+2r^2B'^2(74-81B)+4rB'(B-1)(15B-16)-8\{1-2B^3-4B+5B^2\}\Big)\Big]\Big\}=0,\nonumber\\
& & I= \frac{-3}{2r^{10}\sqrt{R^9}}\Big\{r^6\sqrt{R^7}[r^4BB''''+r^3B'''(rB'+6B)+2r^2B''(B+2rB')-2r^2B'^2+4rB'(1-2B)+4B(B-1)]+r^4\sqrt{R^5}\Big[r^6RBB''''\nonumber\\
& &+r^6BB'''^2-r^3B'''\{r^2B''[rB'-2B]+4r^2B'^2+2rB'(15B-1)+20B(B-1)\}
+2/3r^6B''^3+2r^4B''^2(6rB'-5B-2)\nonumber\\
& &+2r^2B''\{23r^2B'^2+2rB'(14B-9)+4(6B^2-7B+1)\}+104/3r^3B'^3+4r^2B'^2(6B-11)+8rB'(B-1)(2B-3)-8/3(B^3-2)\nonumber\\
& &+8B\Big]-r^4R^2\Big[B\sqrt{R}(4+r^3B'''-4B+4r^2B''-2rB')^2-\alpha\Big(r^6RBB''''+3/2r^6BB'''^2-
r^3B'''\Big\{r^2B''(rB'-6B)+4r^2B'^2\nonumber\\
& &+2rB'(16B-1)+24B(B-1)\Big\}+2r^6B''^3+2r^4B''^2\{10rB'+17B-6\}+2r^2B''\{41r^2B'^2+2rB'(16B-23)+4(3-4B^2+B)\}\nonumber\\
& &+136r^3B'^3-2r^2B'^2(106-117B)+8rB'(B-1)(15B-13)+16(2B-1)(B-1)^2\Big)\Big]\Big\}=0,
\end{eqnarray}
where $q$ is the  gauge potential which is defined as
\begin{equation}  \label{p} A :=q(r)dt.\end{equation}
If we subtract  $I_t{}^t$ from $I_r{}^r$ and solving the system $I_t{}^t-I_r{}^r$ and $I_\theta{}^\theta$, which is a closed system for the two  unknown functions $B(r)$ and $q(r)$,
 we get the following exact solution
\begin{eqnarray} \label{sol}
& &  B(r)=\frac{1}{2}-\frac{1}{3\alpha r}+\frac{1}{3\alpha r^2}, \qquad \qquad A=\frac{1}{\sqrt{3\alpha} r}.\nonumber\\
\end{eqnarray}
The analytic solution (\ref{sol}) satisfy the system of differential equations (\ref{df1}) including the trace equation $I$. Using Eq. (\ref{r1}) we get the Ricci scalar in the form
\begin{equation} \label{ri}
R=\frac{1}{r^2}\,,\end{equation}
which is also a consistency check for the whole procedure.
The metric of the above solution  takes the form
\begin{eqnarray} \label{met5}
& &  ds^2=\left(\frac{1}{2}-\frac{1}{3\alpha r}+\frac{1}{3\alpha r^2}\right)dt^2-\left(\frac{1}{2}-\frac{1}{3\alpha r}+\frac{1}{3\alpha r^2}\right)^{-1}dr^2-r^2d\Omega^2,
\end{eqnarray}
which   asymptotically behaves as a flat space-time. We have to  stress  the fact that solution (\ref{sol}) is different from that obtained in \cite{Sebastiani:2010kv} due to the fact that they derived their solution using the form $f(R)=R+\alpha\sqrt{R}$. Therefore, our solution is identical with their one when we neglect the term ${\displaystyle \frac{1}{3\alpha r^2}}$, which is responsible for the electric charge, and reverse the negative sign to be positive to satisfy the field equation of $f(R)=R+\alpha\sqrt{R}$. We must stress on the fact that the dimensional parameter $\alpha$ must take a positive value so that solution (\ref{sol}) satisfies the field equations (\ref{f1}),  (\ref{f2}) and (\ref{f3}).

%%%%%%%%%%%%%%%%%%%%%%%%%%%%%%%%%%% Section 3 %%%%%%%%%%%%%%%%%%%%%%%%%%%%%%%%%%%%%%%%
\section{An exact (A)dS charged black hole solution  }\label{S4}
%%%%%%%%%%%%%%%%%%%%%%%%%%%%%%%%%%%%%%%%%%%%%%%%%%%%%%%%%%%%%%%%%%%%%%%%%%%%%%%%%%%%%%
Let us derive now a  charged (A)dS black hole solution for the model  $f(R)=R-2\alpha\sqrt{R-8\Lambda}$\footnote{We define  ${\cal R}=R-8\Lambda$.}. Applying the anzatz (\ref{met}) to the field Eqs. (\ref{f1}), (\ref{f2}) and (\ref{f3}), after using (\ref{r1}) and putting $s=1$, we get the following non-vanishing field equations
\newpage
\begin{eqnarray} \label{df2}
& & I_t{}^t=\frac{1}{2r^{10}\sqrt{{\cal {R}}}^9}\Big\{r^6\sqrt{{\cal {R}}^7}[r^4BB''''+B'''(1/2r^4B'+6r^3B)+2r^2B''(B+rB')-r^2B'^2+2rB'(1-3B)+4B(B-1)]\nonumber\\
& & +r^4\sqrt{\cal{ R}^5}\Big[4r^2B'^2+r^6 {\cal{ R}}BB''''-r^6BB'''^2+1/2r^3B'''\{r^2B''(rB'-4B)
+2B'(29rB-r+4r^2\Lambda)+8B(5B+12r^2\Lambda-5)\}\nonumber\\
& &+4r^4B''^2(r^2q''+2r^2\Lambda-6B+2rB'-1)+2r^2B''\Big[23r^2B'^2+2rB'(23B+8r^2q'^2-13+40r^2\Lambda)+8r^2q'^2(B-1+4r^2\Lambda)+48B^2\nonumber\\
& &+8B(8r^2\Lambda-7)+16r^2\Lambda(4r^2\Lambda-3)+8\Big]+28r^3B'^3+
r^2B'^2(34B+32r^2q'^2+184r^2\Lambda-54)+4rB'(B-1+4r^2\Lambda)\Big(7B+8r^2q'^2\nonumber\\
& &-9+24r^2\Lambda\Big)+8r^2q'^2(B-1+4r^2\Lambda)^2+8B^2[14r^2\Lambda-1]+16B(16r^4\Lambda^2-12r^2\Lambda+1)
+8(2r^2\Lambda-1)(4r^2\Lambda-1)^2\Big]\nonumber\\
& &+r^4{\cal R}^2\Big[B\sqrt{{\cal R}}[4r^2B''+r^3B'''-2rB'+4(1-B)]^2+\alpha\Big\{r^6{\cal R}BB''''-
3/2r^6BB'''^2+1/2r^3B'''\Big[r^2B''(rB'-12B)+4r^2B'^2\nonumber\\
& &+2rB'(31B-1+4r\Lambda)+48B(B-1+2r^2\Lambda)\Big]-r^6B''^3-100r^3B'^3+2r^2B'^2[96B-85+324r^2\Lambda]+2r^4B''^2\Big(4-6rB'-15rB\nonumber\\
& &
+16r^2\Lambda\Big)+r^2B''[57r^2B'^2+2rB'[21B-35+68r^2\Lambda]
+4B(4-3B-4r^2\Lambda)+(4r^2\Lambda-1)^2]-4rB'\Big[27B^2+4B(47r^2\Lambda-50)-23\nonumber\\
& &-4r^2\Lambda(88r^2\Lambda-45)\Big]-16[(B-1)^2(2B-1)+2B^2r^2\Lambda+5Br^2\Lambda(3r^2\Lambda-11)+(4r^2\Lambda-1)^3]\Big\}\Big]\Big\}=0,\nonumber\\
& & I_r{}^r=- \frac{1}{4r^8\sqrt{{\cal R}^7}}\Big\{r^4\sqrt{{\cal R}^5}\Big[(rB'+4B)[4(1-B)-2rB'+r^2(4B''+rB''')]-\Big(r^3B'''[rB'+4B]+4r^2B''[5B+r^2q'^2+2rB'-1\nonumber\\
& &
+2r^2\Lambda]+14r^2B'^2+rB'[16r^2q'^2+12B-20+64r^2\Lambda]+8(B-1+8r^2\Lambda)r^2q'^2+8+64r^4\Lambda^2-8B^2-48r^2\Lambda(1+B)\Big)\Big]\nonumber\\
& &-\alpha r^4{\cal R}^2\Big(r^3B'''[4B+rB']-2r^4B''^2+r^2B''[4(B+3)+48r^2\Lambda-16rB']-50r^2B'^2+4rB'(15-17B-r^2\Lambda)-16[2B-1][B-1]\nonumber\\
& &+32r^2\Lambda(4B+8r^2\Lambda+1)\Big)\Big\}=0,\nonumber\\
 & & I_\theta{}^\theta=I_\phi{}^\phi=
 \frac{-1}{2r^{10}\sqrt{{\cal R}^9}}\Big\{r^6\sqrt{{\cal R}^7}\Big[r^4BB''''+r^3B'''(rB'+5B)+2r^2B''(2rB'-B)+2rB'(2-3B)-2r^2B'^2+8B(B-1)\Big]\nonumber\\
& &-r^5\sqrt{{\cal R}^5}\Big[r^5BRB''''-r^5BB'''^2
 +r^2B'''\Big\{r^2B''(rB'-3B)+4r^2B'^2+2rB'(13B-1+2r^2\Lambda)+2B(9B-9+20r^2\Lambda)\Big\}+r^5B''^3\nonumber\\
& &-2r^3B''^2(r^2q'^2+2-7rB'+7B-10r^2\Lambda)\Big]+2rB''\Big(23r^2B'^2-rB'[14+8r^2q'^2-17B-80r^2\Lambda]+4(B-1-4r^2\Lambda)r^2q'^2\nonumber\\
& &-B[10B+8r^2\Lambda-11]+4r^2\Lambda[3+8r^2\Lambda]+1\Big)
 +24r^2B'^3-4rB'^2(3+8r^2q'^2-44r^2\Lambda)+4B'\Big[(B-1+4r^2\Lambda)r^2q'^2\nonumber\\
& &+B[3B-3+20r^2\Lambda]+24r^2\Lambda[4r^2\Lambda-1]-r[q'^2(B-1+4r^2\Lambda)-2B(5B+4r^2\Lambda-3)+4(r^2\Lambda-1)^2]\Big]+r^4{\cal R}^2\Big[\sqrt{{\cal R}}B\Big[B-1\nonumber\\
& &
-r^3B'''-4r^2B''+2rB'\Big]^2+\alpha\Big(r^6{\cal R} BB''''+3/2r^6BB'''^2 -r^3B'''\Big[r^2B''\{rB'-7B\}+4r^2B'^2+2rB'[14B-1+4r^2\Lambda]\nonumber\\
& &+2B(11B-11+20r^2\Lambda)\Big]+2r^6B''^3+2r^4B''^2[18B+9rB'-5+24r^2\Lambda]+2r^2B''\Big[33r^2B'^2+rB'(27B-34+160r^2\Lambda)\nonumber\\
& &-2(9B^2-5B-4)-r^2\Lambda\{88B+192r^2\Lambda+80\}\Big]
 +104r^3B'^3+2r^2B'^2(74-81B+328r^2\Lambda)+rB'\Big[15B^2-31B\nonumber\\
& &+156Br^2\Lambda+352r^4\Lambda^2+16-152r^2\Lambda\Big]-8\{1-2B^3-8r^2\Lambda-4B+5B^2-90r^4\Lambda^2+24r^2\Lambda\}+(8r^2\Lambda-1)
(4r^2\Lambda-1)^2\Big)\Big]\Big\}=0,\nonumber\\
& & I= \frac{-3}{2r^{10}\sqrt{{\cal R}^9}}\Big\{r^6\sqrt{{\cal R}^7}[r^4BB''''+r^3B'''(rB'+6B)+2r^2B''(B+2rB')-2r^2B'^2+4rB'(1-2B)+4B(B-1)]+r^4\sqrt{{\cal R}^5}\Big[r^6{\cal R}BB''''\nonumber\\
& &+r^6BB'''^2-r^3B'''\{r^2B''[rB'-2B]+4r^2B'^2+2rB'(15B-1+4r^2\Lambda)+4B(5B-5+12r^2\Lambda)\}
+2/3r^6B''^3+2r^4B''^2(6rB'-5B\nonumber\\
& &-2+8r^2\Lambda)+2r^2B''\{23r^2B'^2+2rB'(14B-9+40r^2\Lambda)+4B(6B-7+10r^2\Lambda)+(4r^2\Lambda-1)^2\}+104/3r^3B'^3+4r^2B'^2(6B-11\nonumber\\
& &+60r^2\Lambda)+8rB'(B-1+4r^2\Lambda)(2B-3+16r^2\Lambda)-8/3B^3+8B+96B^2r^2\Lambda+32r^2B\Lambda[8r^2\Lambda-5]+64/3(4r^2\Lambda-1)^3\Big]-r^4{\cal R}^2\nonumber\\
& &\Big[B\sqrt{{\cal R}}(4+r^3B'''-4B+4r^2B''-2rB')^2-\alpha\Big(r^6{\cal R}BB''''+3/2r^6BB'''^2-
r^3B'''\Big\{r^2B''(rB'-6B)+4r^2B'^2+2rB'(16B-1\nonumber\\
& &+4r^2\Lambda)+24B(B-1+2r^2\Lambda)\Big\}+2r^6B''^3+2r^4B''^2\{10rB'+17B-6+80/3r^2\Lambda\}+2r^2B''\{41r^2B'^2+2rB'(16B-23+296/3r^2\Lambda)\nonumber\\
& &+4(3-4B^2+B+176/3r^2\Lambda)+16/3r^2\Lambda(r^2\Lambda-20)\}+136r^3B'^3-2r^2B'^2(106-117B-1304/3r^2\Lambda)+8rB'(15B^2+B[344/3r^2\Lambda\nonumber\\
& &-84]+r^2\Lambda/3[704r^2\Lambda-332]+13)+32B^3+16B^2[34/3r^2\Lambda-5]+32B[2-37/3r^2\Lambda+88/3r^4\Lambda^2]+16(4r^2\Lambda-1)^2\nonumber\\
& &
\times(16r^2\Lambda-3)\Big)\Big]\Big\}=0.\nonumber\\
& &
\end{eqnarray}
If we subtract the component $I_t{}^t$ from the component $I_r{}^r$ and solve the system $I_t{}^t-I_r{}^r$ and $I_\theta{}^\theta$ which is a closed system for the two unknowns functions $B(r)$ and $q(r)$,
 we get the  exact solution
\begin{eqnarray} \label{sol1}
& &  B(r)=\frac{1}{2}-\frac{2r^2\Lambda}{3}-\frac{1}{3\alpha r}+\frac{1}{3\alpha r^2}, \qquad \qquad A=\frac{1}{\sqrt{3\alpha} r},\nonumber\\
\end{eqnarray}
Using Eq. (\ref{sol1}) in (\ref{r1}) we get the Ricci scalar in the form
\begin{equation} \label{ri}
R=\frac{8r^2\Lambda+1}{r^2}.\end{equation}
The metric of the above solution takes the form
\begin{eqnarray} \label{met3}
& &  ds^2=\left(\frac{1}{2}-\frac{2r^2\Lambda}{3}-\frac{1}{3\alpha r}+\frac{1}{3\alpha r^2}\right)dt^2-\left(\frac{1}{2}-\frac{2r^2\Lambda}{3}-\frac{1}{3\alpha r}+\frac{1}{3\alpha r^2}\right)^{-1}dr^2-r^2d\Omega^2.
\end{eqnarray}
which behaves asymptotically as (A)dS spacetime. Solution (\ref{sol1}) is different from that derived in \cite{Sebastiani:2010kv} due to the reason discussed for solution (\ref{sol}).    Same constrain put on  the parameter $\alpha$ in the non charge case is also true here. 

\section{Physical properties of the black holes }\label{S55}

The metric of solution (\ref{sol})  can be rewritten in  the form
\begin{equation} \label{me}
ds^2=\left(\frac{1}{2}-\frac{2M}{r}+\frac{q^2}{r^2}\right)dt^2-\left(\frac{1}{2}-\frac{2M}{r}+\frac{q^2}{r^2}\right)^{-1}dr^2-r^2d\Omega^2\;, \qquad \textrm{where} \qquad M=\frac{1}{6\alpha}, \qquad q=\frac{1}{\sqrt{3\alpha}},\end{equation} which shows clearly that the dimensional parameter $\alpha$ cannot be equal zero and, in that case, the line element coincides with the  Reissner-Nordstr\"om spacetime. Also the metric of solution (\ref{sol1})  may be rewritten as
\begin{equation} \label{me1}
ds^2=\left(\frac{1}{2}-\frac{2r^2\Lambda}{3}-\frac{2M}{r}+\frac{q^2}{r^2}\right)dt^2-\left(\frac{1}{2}-\frac{2r^2\Lambda}{3}-\frac{2M}{r}+\frac{q^2}{r^2}\right)^{-1}dr^2-r^2d\Omega^2\;, \qquad \textrm{where, again} \qquad M=\frac{1}{6\alpha} \qquad \textrm{and} \qquad q=\frac{1}{\sqrt{3\alpha}}.\end{equation} which shows   that  line element coincides with the  (A)dS Reissner-Nordstr\"om spacetime.   Equations (\ref{me}) and (\ref{me1}) show in a clear way that the dimensional parameter $\alpha$ must not equal zero.

Let us  study now the regularity of the solutions (\ref{sol}) and (\ref{sol1}) when $B(r)=0$. For   solution  (\ref{sol}), we evaluate the scalar invariants and get
\begin{eqnarray} \label{scal1}
&&R^{\mu \nu \lambda \rho}R_{\mu \nu \lambda \rho}= \frac{56+9r^4\alpha^2+12\alpha r^3+12r^2[\alpha+1]+48r}{9\alpha^2r^8}, \nonumber\\
& &R^{\mu \nu}R_{\mu \nu}=\frac{9r^4\alpha^2-12\alpha r^2+8}{18\alpha^2r^8}, \qquad \qquad R= \frac{1}{r^2},
\end{eqnarray}
where $R^{\mu \nu \lambda \rho}R_{\mu \nu \lambda \rho}$, $R^{\mu \nu}R_{\mu \nu }$, $R$ are the Kretschmann scalars,  the Ricci tensor  square, the Ricci scalar, respectively.   Equations  (\ref{scal1}) show that the solutions, at $r=0$, have true singularities and the dimensional parameter  $\alpha\neq0$. Also Eq. (\ref{sol}) as well as Eq. (\ref{met5})   show  clearly that the dimensional parameter $\alpha$  cannot be equal to zero which insure that solution (\ref{sol}) cannot reduce to GR. This means that this solution  is a new exact charged one in the frame of $f(R)$ gravitational theory.

Using Eq. (\ref{sol1})  we get the scalar invariants in the form
\begin{eqnarray} &&R^{\mu \nu \lambda \rho}R_{\mu \nu \lambda \rho} =\frac{96r^8\Lambda^2 \alpha^2+24r^6\Lambda\alpha^2+9r^4 \alpha^2+12\alpha r^3+12r^2[\alpha+1]+48r+56}{9\alpha^2r^8},
 \nonumber\\
&&R^{\mu \nu}R_{\mu \nu} =\frac{288r^8\Lambda^2 \alpha^2+72r^6\Lambda \alpha^2+9\alpha^2 r^4-12\alpha r^2+8}{18 \alpha^2 r^8},\qquad \qquad \qquad
R =\frac{8r^2\Lambda+1}{r^2}.\nonumber\\
\end{eqnarray}
The same considerations carried out for solution (\ref{sol}) can also be applied for solution (\ref{sol1}) which insure also that solution (\ref{sol1}) is a novel charged one in the framework  of $f(R)$ gravity that cannot reduce to GR.
 %%%%%%%%%%%%%%%%%%%%%%%%%%%% Section 7 %%%%%%%%%%%%%%%%%%%%%%%%%%%%%
\section{Black hole thermodynamics }\label{S66}
%%%%%%%%%%%%%%%%%%%%%%%%%%%%%%%%%%%%%%%%%%%%%%%%%%%%%%%%%%%%%%%%%%%%
Now we are going to explore the thermodynamics of the new  black hole solutions derived in the previous sections. The Hawking temperature is defined as \cite{PhysRevD.86.024013,Sheykhi:2010zz,Hendi:2010gq,PhysRevD.81.084040}
  \begin{equation}
T_+ = \frac{B'(r_+)}{4\pi},
\end{equation}
where the event horizon is located at $r = r_+$ which is the largest positive root of $B(r_+) = 0$ that fullfils $B'(r_+)\neq 0$.
The Bekenstein-Hawking entropy  in the framework of $f(R)$ gravity is given as \cite{PhysRevD.84.023515,PhysRevD.86.024013,Sheykhi:2010zz,Hendi:2010gq,PhysRevD.81.084040,Zheng:2018fyn}
\begin{equation}\label{ent}
S(r_+)=\frac{1}{4}Af_{R}(r_+),
\end{equation}
where $A$  is the area of the event horizon. The form of the quasi-local  energy in the framework of $f(R)$ gravity is defined as \cite{PhysRevD.84.023515,PhysRevD.86.024013,Sheykhi:2010zz,Hendi:2010gq,PhysRevD.81.084040,Zheng:2018fyn}
\begin{equation}\label{en}
E(r_+)=\frac{1}{4}\displaystyle{\int }\Bigg[2f_{R}(r_+)+r_+{}^2\Big\{f(R(r_+))-R(r_+)f_{R}(r_+)\Big\}\Bigg]dr_+.
\end{equation}

At the horizon, one has the constraint $B(r_+) = 0$ which gives
\begin{eqnarray} \label{m33}
&& {r_+}_{{}_{{}_{{}_{{}_{\tiny Eq. (\ref{sol})}}}}}=\frac{1}{3\alpha}\left[1+\sqrt{1+6\alpha}\right], \qquad \qquad {r_-}_{{}_{{}_{{}_{{}_{\tiny Eq. (\ref{sol})}}}}}=\frac{1}{3\alpha}\left[1- \sqrt{1+6\alpha}\right]
 \nonumber\\
&&{r_+}_{{}_{{}_{{}_{{}_{\tiny Eq. (\ref{sol1})}}}}}=Root(4x^4\alpha \Lambda-3\alpha x^2+2x+2),
\end{eqnarray}
where $Root(4x^4\alpha \Lambda-3\alpha x^2+2x+2)$ is the roots of the equation $(4x^4\alpha \Lambda-3\alpha x^2+2x+2=0)$.  It is clear from the first equation of  Eqs. (\ref{m33})  that $\alpha$ should not be equal zero to ensures that the black hole (\ref{sol}) has no analogy with GR. Moreover,  Eqs. (\ref{m33})   tell us  that the dimensional parameter  $\alpha$ should be positive so that the horizons have a positive real value. Therefore we must put the  restriction  $\alpha>0$, otherwise we get a non-real value for the horizon.  This constraint is consistent with the relation given by Eqs. (\ref{me}) and (\ref{me1}) which allows  the mass parameter to have the correct sign in the metric and  the charge parameter has a real value. Moreover, if the parameter $\alpha$ takes a negative value,   the solutions (\ref{sol}) and (\ref{sol1}) do not satisfy the field Eqs. (\ref{f1}), (\ref{f2}) and (\ref{f3}).

\begin{figure}
\centering
  \includegraphics[scale=0.4]{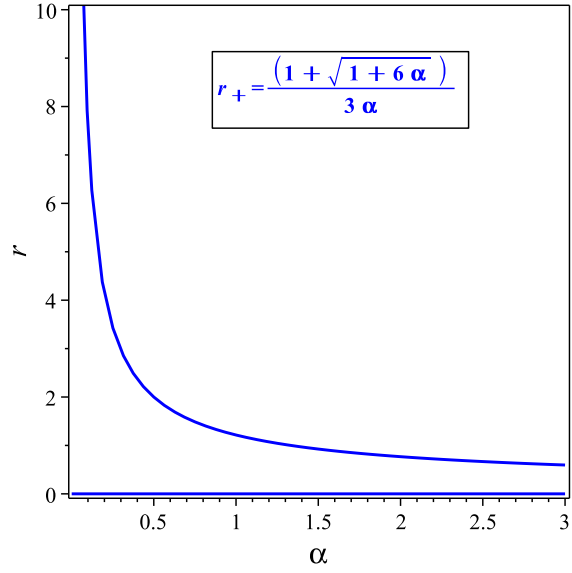}\hspace{0.2cm}
%\subfigure[~The spherically symmetric AdS/dS spacetime]{\label{fig:1b}\includegraphics[scale=0.4]{JMTh2}}
\caption{Schematic plot of the radial coordinate  $r$  versus the dimensional parameter $\alpha$ that characterize the  spherically symmetric black hole [12].}
\label{Fig:1}
\end{figure}

 The  relation between the radial coordinate $r$ and the dimensional parameter $\alpha$ of the black hole (\ref{sol}) is represented in Figure \ref{Fig:1}. From this figure  we can see the
 root of $B(r)$  defining the black hole  outer event  horizon $r_+$ \cite{Brecher:2004gn}. We can continue the study  of  thermodynamics assuming $\alpha>0$ according to the pervious analysis and taking into account the outer event  horizon $r_+$ only which is consistent with  $\alpha>0$.

\begin{figure}
\centering
\subfigure[~The entropy of the black hole solution (\ref{sol})  ]{\label{fig:2a}\includegraphics[scale=0.4]{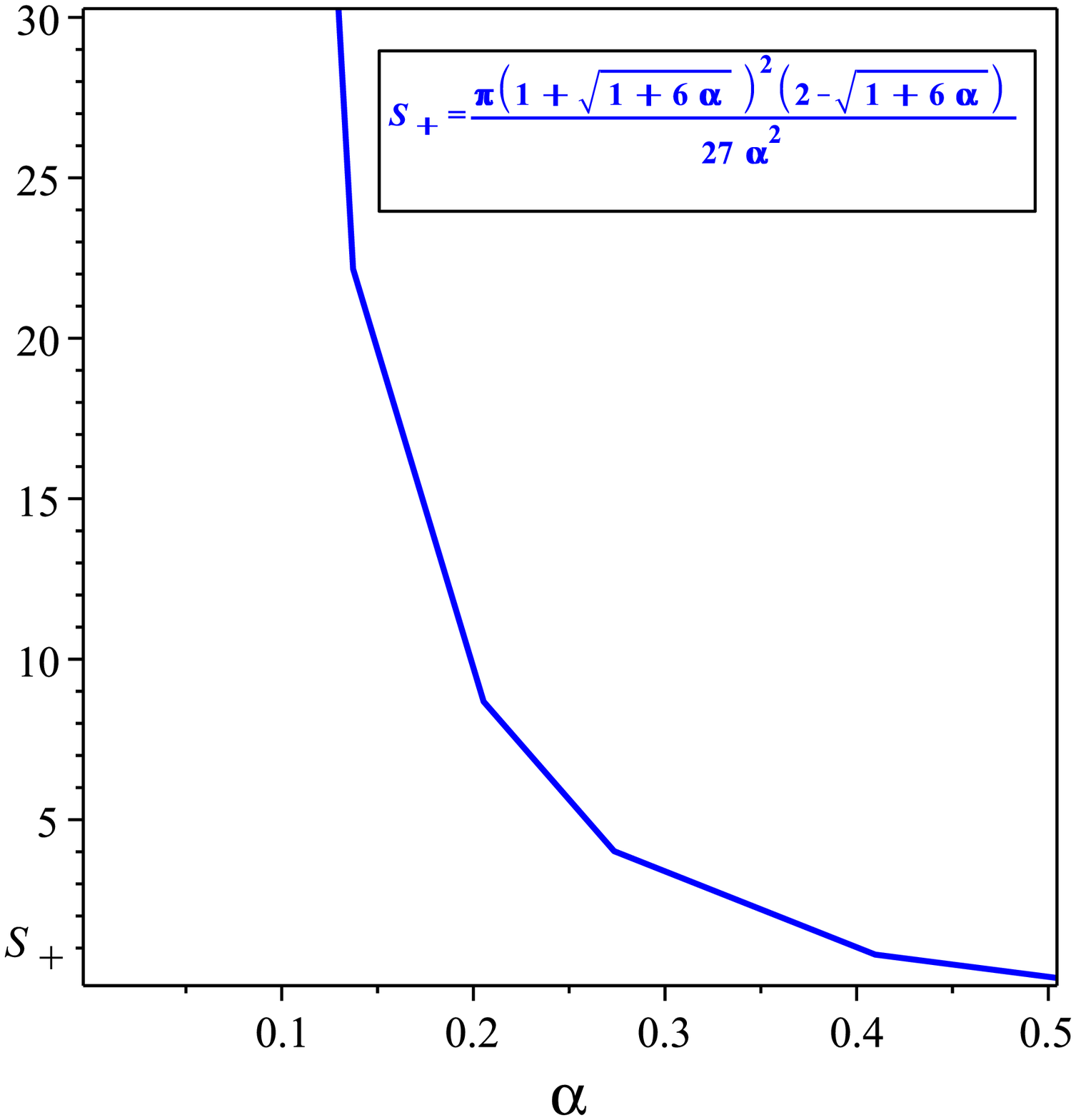}}\hspace{0.2cm}
%\subfigure[~The entropy of the black hole solution (\ref{sol})  ]{\label{fig:2a}\includegraphics[scale=0.4]{JMTh3rr}}\hspace{0.2cm}
\subfigure[~The entropy of the black hole solution (\ref{sol1})]{\label{fig:2b}\includegraphics[scale=0.4]{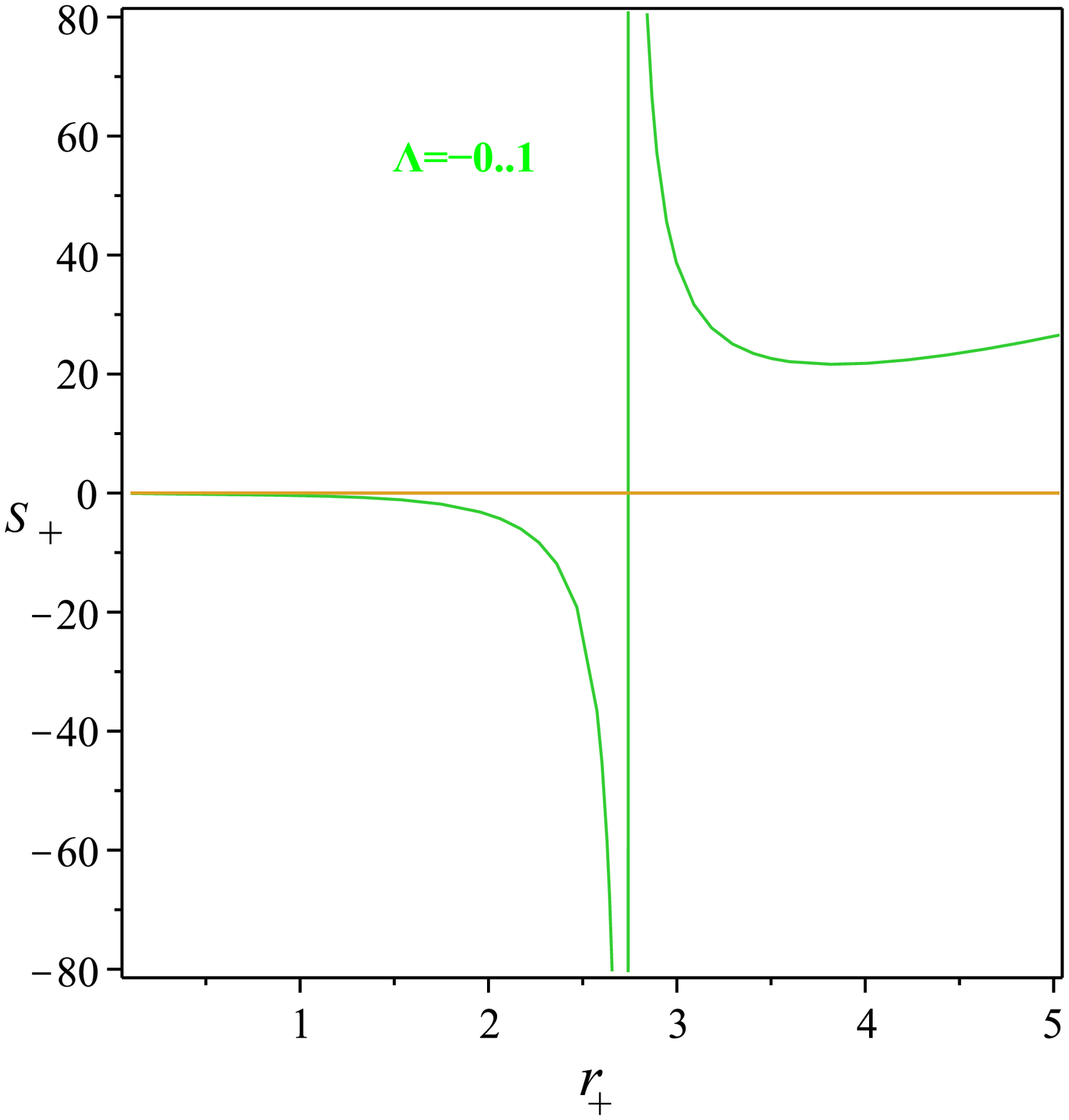}}
\caption{ Schematic plot of the entropy  of the two black holes (\ref{sol}) and (\ref{sol1}) versus the dimensional parameter $\alpha$ and $r_+$ respectively.}
\label{Fig:2}
\end{figure}

\begin{figure}
\centering
\subfigure[~The Hawking temperature of the black hole solution (\ref{sol})  ]{\label{fig:3a}\includegraphics[scale=0.4]{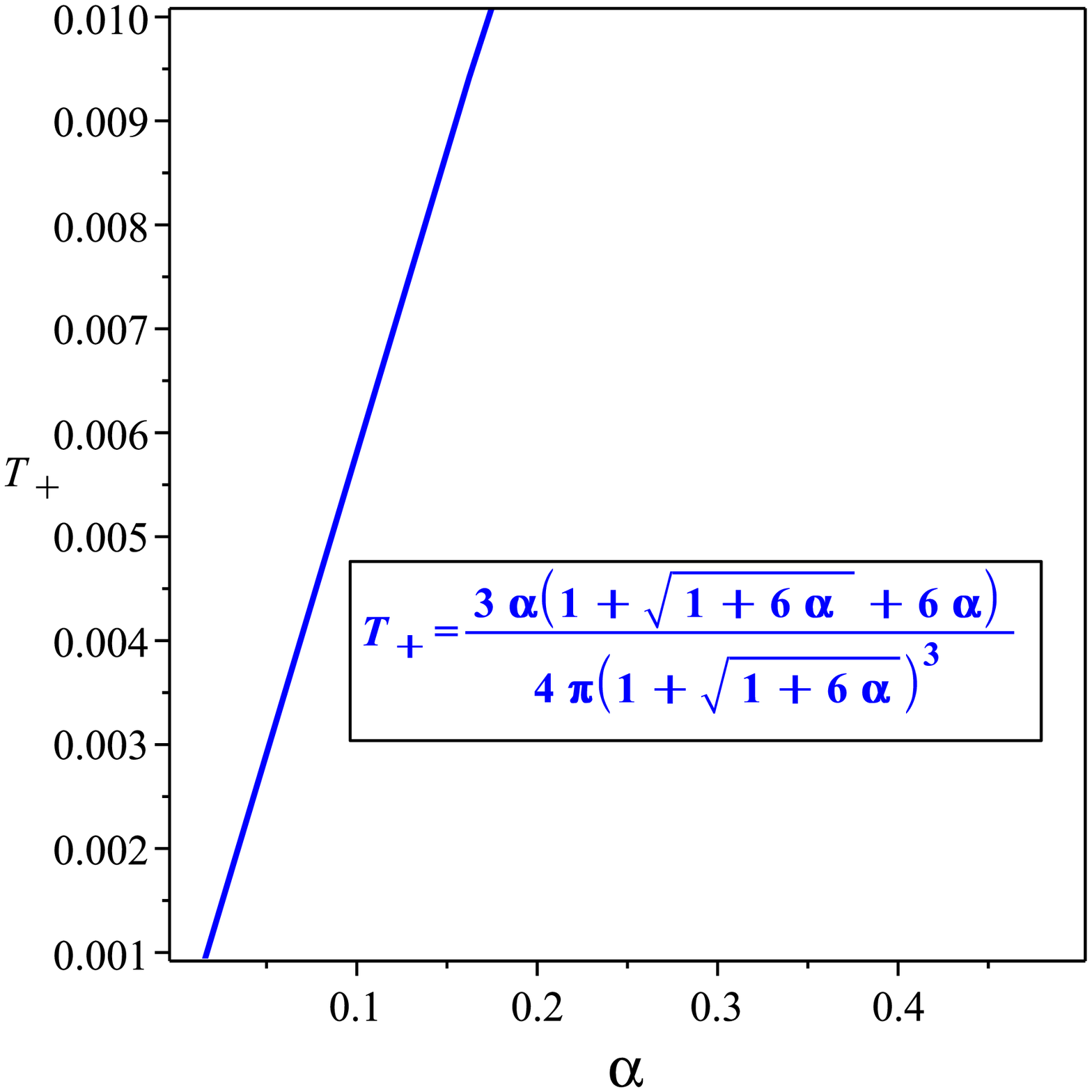}}\hspace{0.2cm}
\subfigure[~The Hawking temperature of the black hole solution (\ref{sol1})]{\label{fig:3b}\includegraphics[scale=0.4]{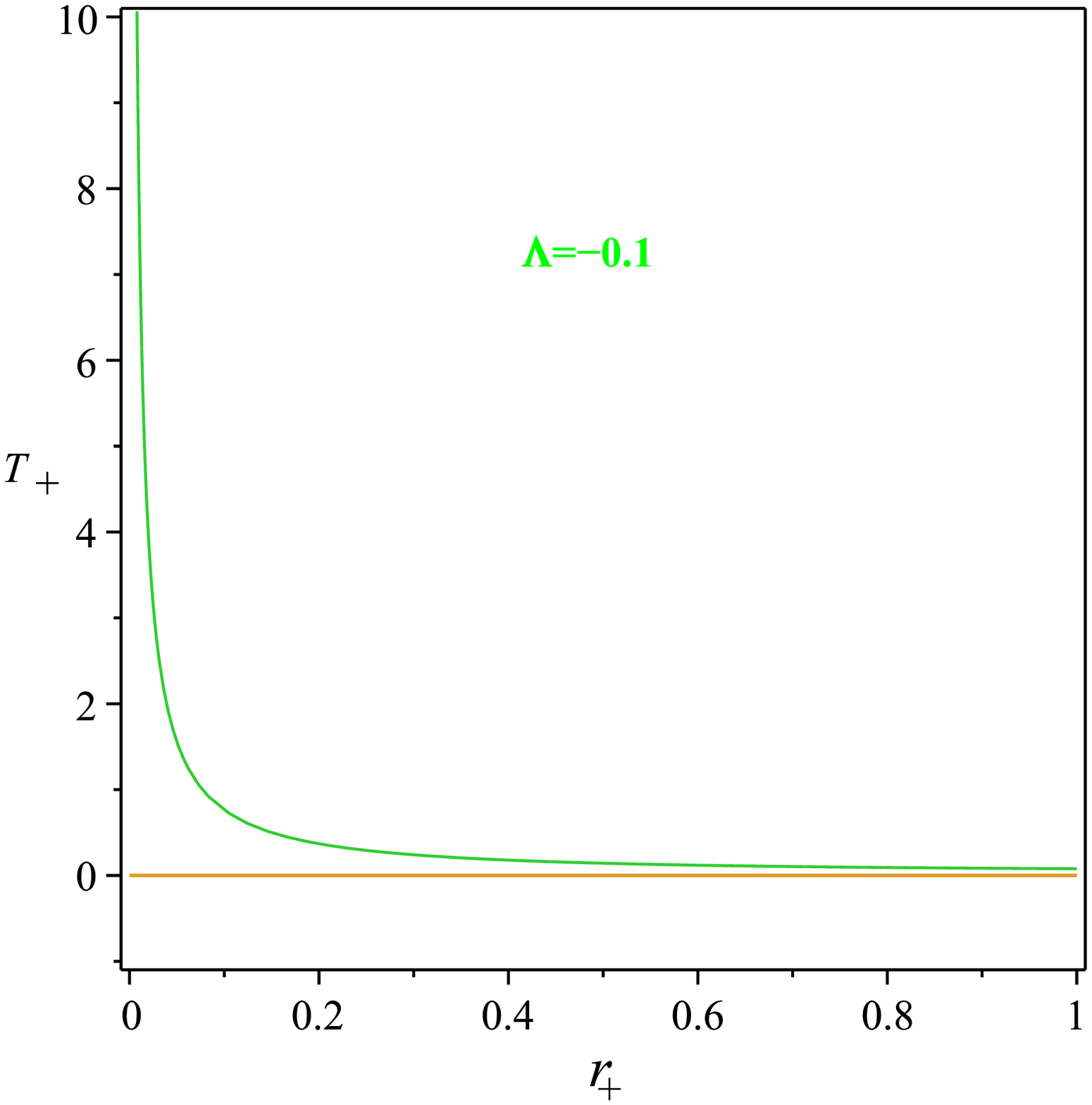}}
\caption{Schematic plot of the Hawking temperature  of the two black holes (\ref{sol}) and (\ref{sol1}) versus the dimensional parameter $\alpha$ and $r_+$ , respectively.}
\label{Fig:3}
\end{figure}

Using Eq. (\ref{ent}), the entropy of the black holes (\ref{sol}) and (\ref{sol1}) are computed as
\begin{eqnarray} \label{ent1}
{S_+}_{{}_{{}_{{}_{{}_{\tiny Eq. (\ref{sol})}}}}}&=&\frac{\pi}{27\alpha^2}\left[1+ \sqrt{1+6\alpha}\right]^2\left[2-\sqrt{1+6\alpha}\right], \nonumber\\
{S_+}_{{}_{{}_{{}_{{}_{\tiny Eq. (\ref{sol1})}}}}}&=&\frac{\pi  r_+{}^2}{4}\left[1-\alpha r_+\right].
\end{eqnarray}
 The first equation  of Eq. (\ref{ent1}) shows that, in order to have  a positive entropy, the dimensional parameter $\alpha$ must take the value $0<\alpha<0.5$.  The second equation of (\ref{ent1}) tells us that  we must have $\alpha<\frac{1}{r_+}$ for positive entropy. Eqs. (\ref{ent1}) are drawn in Figure \ref{Fig:2}. As one can see, from   Figure \ref{Fig:2} \subref{fig:2a},  for $0.5>\alpha>0$ the black hole (\ref{sol}) has $+ve$ entropy. For the black hole\footnote{We substitute the value of $\alpha$ in terms of $\Lambda$ using Eq. (\ref{sol1}) through  this section.} (\ref{sol1}), as Figure \ref{Fig:2} \subref{fig:2b} shows, we have a  phase transition at $2.738612788$ then the entropy has a $-ve$ value  for $0<\alpha<2.738612788$ and it evolves to $+ve$   at $r=2.187$.  The following remarks must be taken into account: It is remarkable that the entropy $S$ is not proportional to the area of the horizon due to Eq. (\ref{ent}). We should also note that the entropy $S$ is proportional to the area if there is no Ricci scalar squared  term i.e. $f_R=1$.
The Hawking temperatures associated with the black hole solutions (\ref{sol}) and (\ref{sol1}) are
\begin{eqnarray} \label{m44}
{T_+}_{{}_{{}_{{}_{{}_{\tiny Eq. (\ref{sol})}}}}}&=&\frac{3\alpha(1+\sqrt{1+6\alpha}+6\alpha)}{4\pi(1+\sqrt{1+6\alpha})^3},\nonumber\\
{T_+}_{{}_{{}_{{}_{{}_{\tiny Eq. (\ref{sol1})}}}}}&=&\frac{r_+-4\alpha \Lambda r_+{}^4+2}{12\pi \alpha r_+{}^3},
\end{eqnarray}
where ${T_+}$ is the Hawking temperature at the event horizon. We represent the Hawking temperature in Figure \ref{Fig:3}.   Figure  \ref{Fig:3} \subref{fig:3a}, which is related to the black hole (\ref{sol}), shows that we  have a   positive temperature  when the parameter $\alpha$ has the value $0<\alpha<0.5$. Figure  \ref{Fig:3} \subref{fig:3b} is related to the black hole (\ref{sol1}). Here the temperature  has always a  $+ve$ value.
From Eq. (\ref{en}), the quasi-local energy of the two black holes (\ref{sol}) and (\ref{sol1}) are calculated as
\begin{eqnarray} \label{m44}
{E_+}_{{}_{{}_{{}_{{}_{\tiny Eq. (\ref{sol})}}}}}&=&\frac{1+\sqrt{1+6\alpha}-3\alpha}{12 \alpha},\nonumber\\
{E_+}_{{}_{{}_{{}_{{}_{\tiny Eq. (\ref{sol1})}}}}}&=&\frac{r_+}{8}\left(4-3\alpha r_++4\Lambda \alpha r_+{}^3\right).
\end{eqnarray}
\begin{figure}
\centering
\subfigure[~The quasilocal energy of the black hole solution (\ref{sol})  ]{\label{fig:4a}\includegraphics[scale=0.4]{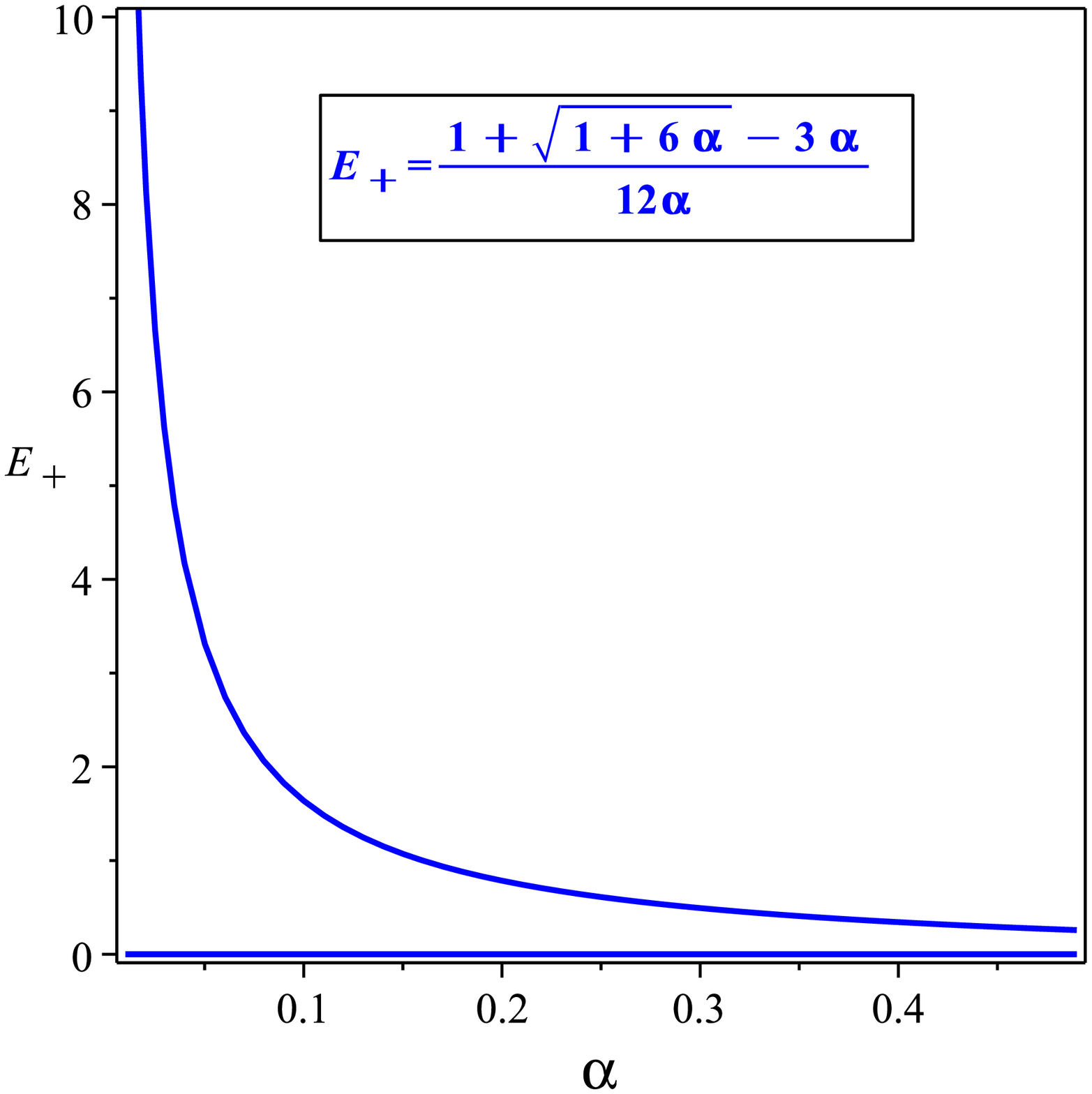}}\hspace{0.2cm}
\subfigure[~The quasilocal energy of the black hole solution (\ref{sol1})]{\label{fig:4b}\includegraphics[scale=0.4]{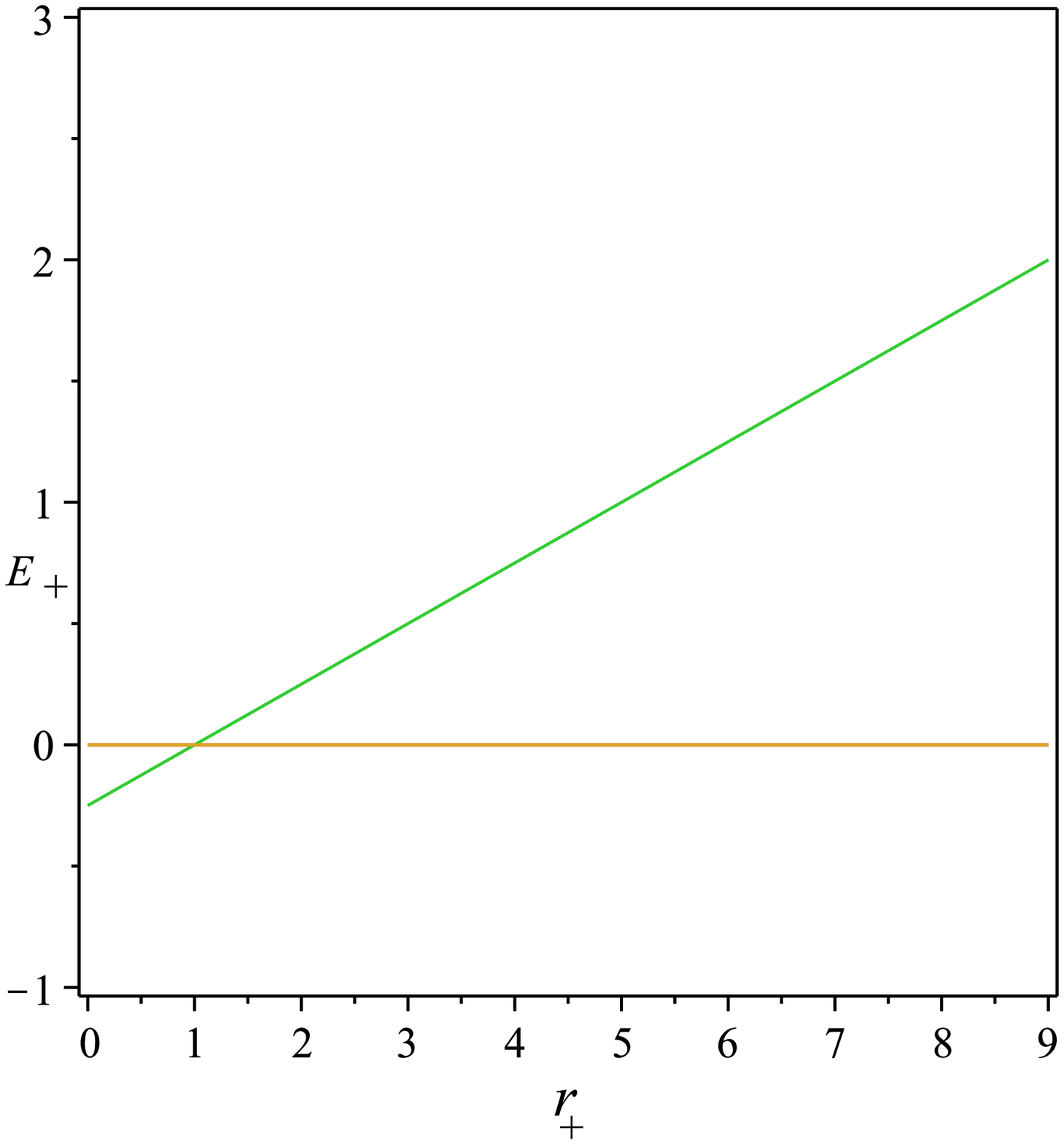}}
\caption{Schematic plot of the quasilocal energy  of the   black holes (\ref{sol}) and    (\ref{sol1})  versus the dimensional parameter $\alpha$ and $r_+$, respectively.}
\label{Fig:4}
\end{figure}
 The first equation of (\ref{m44})  shows that the dimensional parameter has to be  $\alpha\neq 0$. We plot the energy   in Figure \ref{Fig:4} which shows that for Figure  \ref{Fig:4} \subref{fig:4a},  the quasi-local energy   has  a $+ve$ value when $0<\alpha<0.5$.  In the other case, we have a negative value for the quasi-local energy till $r_+=1$ and then the energy becomes positive as  Figure \ref{Fig:4} \subref{fig:4b} shows.

The free energy in the grand canonical ensemble, also called Gibbs free energy, can be defined as \cite{Zheng:2018fyn,Kim:2012cma}
\begin{equation} \label{enr}
G(r_+)=E(r_+)-T(r_+)S(r_+)%+P(r_+)V(r_+),
\end{equation}
%$V$ is the geometric volume  of the black hole, $P$ is the pressure which is represented by the radial    equation of (\ref{f1}), i.e. $I_r{}^r$,
where   $E(r_+)$, $T(r_+)$ and $S(r_+)$ are  the quasilocal energy, the temperature and entropy  at the event horizons, respectively.   Using Eqs. (\ref{ent}), (\ref{m33}), (\ref{ent1}) and (\ref{m44}) in (\ref{enr}) we get
\begin{eqnarray} \label{m77}
&&{G_+}_{{}_{{}_{{}_{{}_{\tiny Eq. (\ref{sol})}}}}}=\frac{1+\sqrt{1+6\alpha}-3\alpha}{12\alpha}-\frac{\alpha r_+{}^2(1+\sqrt{1+6\alpha}+6\alpha)(2-\sqrt{1+6\alpha})}{4(1+\sqrt{1+6\alpha})^3}, \nonumber\\
 &&{G_+}_{{}_{{}_{{}_{{}_{\tiny Eq. (\ref{sol1})}}}}}=\frac{r_+(4-3\alpha r+4\Lambda \alpha r_+{}^3)}{8}-\frac{(r-4\Lambda \alpha r^3+2)(1-\alpha r)}{48\alpha r_+}.
\end{eqnarray}
\begin{figure}
\centering
\subfigure[~The free energy of the black hole solution (\ref{sol})  ]{\label{fig:5a}\includegraphics[scale=0.4]{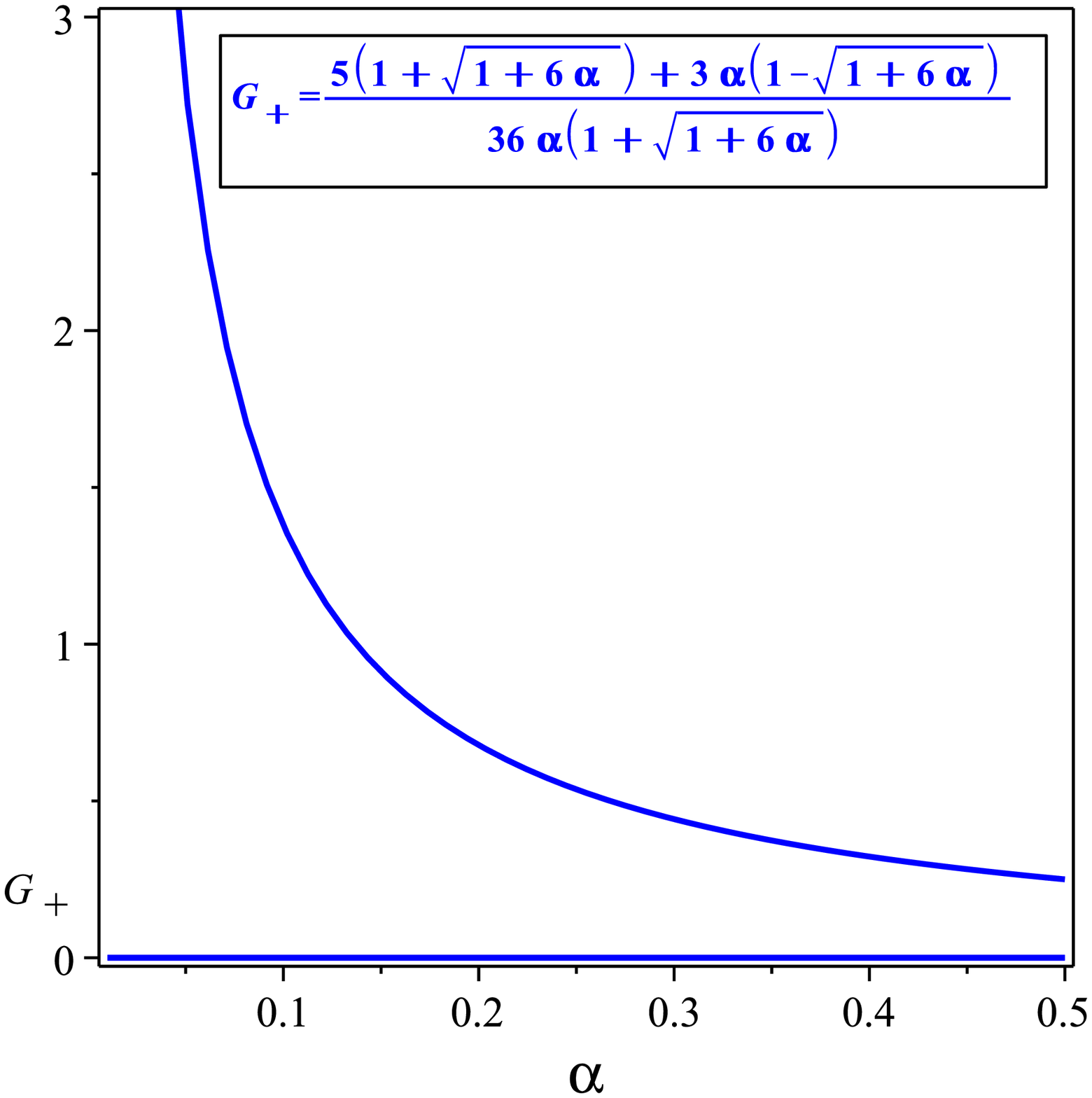}}\hspace{0.2cm}
\subfigure[~The free energy of the black hole solution (\ref{sol1})]{\label{fig:5b}\includegraphics[scale=0.4]{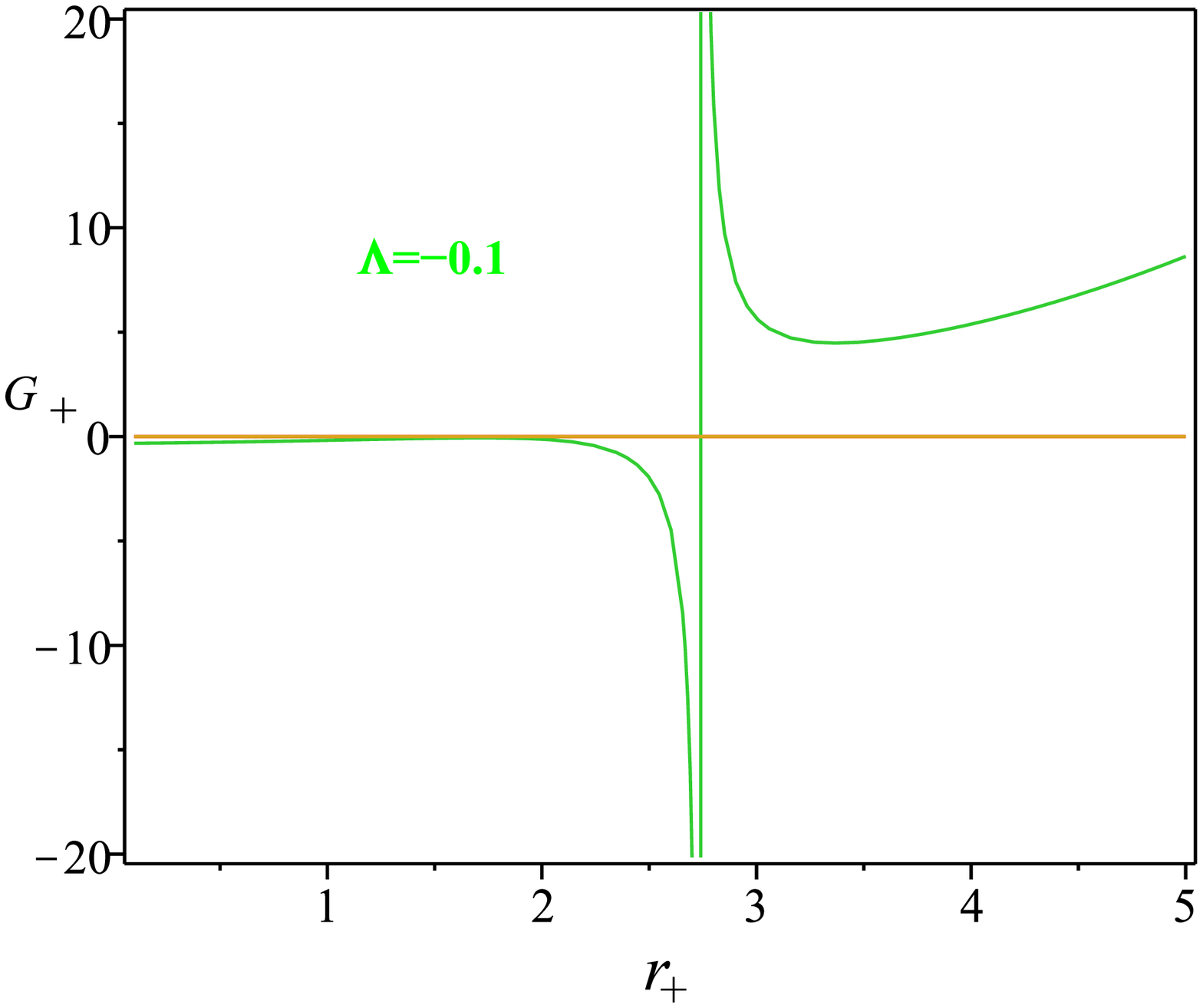}}
\caption{ Schematic plot of the free energy  of the  black holes  (\ref{sol}) and    (\ref{sol1})  versus the dimensional parameter $\alpha$ and $r_+$, respectively.}
\label{Fig:5}
\end{figure}
%where we have put $V(r_+)=\frac{4}{3}\pi r_+{}^3$ and
The behaviors of the Gibbs energy of our black holes are presented in Figures \ref{Fig:5}\subref{fig:5a}, \ref{Fig:5}\subref{fig:5b} for particular values of the model parameters. As Figure \ref{Fig:5}\subref{fig:5a} shows, for the black hole solution (\ref{sol}), the Gibbs energy is positive when $0<\alpha <0.5$ which means that it is more globally stable. For the black hole solution (\ref{sol1}),  the  Gibbs energy has a  phase transition. It has a negative value   when $r<2.73$ and positive when $2.73<r$.   
%%%%%%%%%%%%%%%%%%%%%%%%%%%% Section 7 %%%%%%%%%%%%%%%%%%%%%%%%%%%%%
\section{The stability of  charged black hole solutions  in $f(R)$ gravity}\label{S616}
%%%%%%%%%%%%%%%%%%%%%%%%%%%%%%%%%%%%%%%%%%%%%%%%%%%%%%%%%%%%%%%%%%%%
 In order to study the stability of the above black hole solutions, it is better recast  $f(R)$ gravity in terms of the corresponding scalar- tensor theory.
%The Lagrangian of the gravitational sector is given by
%\begin{equation} \label{action1}
%S=\frac{1}{2\kappa}\int d^{4}x\,\sqrt{-g}\, f(R).
%\end{equation}
Discarding the cosmological term, the Lagrangian (\ref{a2}) can be rewritten
as
\begin{equation}
S=\frac{1}{2\kappa}\int d^{4}x\sqrt{-g}\,[\phi\, R-V(\phi)],\label{action2}
\end{equation}
where $\phi$ is a scalar field   coupled to the Ricci scalar $R$ and $V(\phi)$ is the potential (see \cite{Capozziello:2011et} for details).
Here,  we will discuss the behavior of the perturbations about
a static spherically symmetric  vacuum background, whose metric is  written as above, that is
\begin{equation}
ds^{2}=g_{\mu\nu}^{0}dx^{\mu}dx^{\nu}=B(r)\, dt^{2}-\frac{dr^{2}}{B(r)}- r^{2} ( d\theta^{2} +\sin^2\theta\, d\phi^{2}).
\end{equation}
where $g_{\mu\nu}^{0}$ is the background metric.
Considering the above black hole solutions, we want to investigate  whether these
backgrounds are stable or not against linear perturbations, and what
we can learn in terms of speed of propagation for the
scalar gravitational modes. For such a theory, the background
equations of motion read
\begin{equation}
  V =- {\frac{4B\, \phi'}{r}}-{\frac{2\phi\, B'}{r}}-\phi' B'+{\frac{2\phi}{{r}^{2}}}-{\frac{2B\, \phi}{{r}^{2}}}\, , \qquad \ \qquad \phi'' =0\,,\qquad \qquad   R =\frac{dV}{d\phi}\,,\label{eq:Ueq}\\
\end{equation}
 where $'$ stands for differentiation with respect to $r$.

\subsection{Outline  of the Regge-Wheeler-Zerilli formalism}

Before studying the metric perturbation of  static spherically
symmetric  spacetime of $f(R)$ gravity, let us  give a brief
summary of the formalism developed by Regge, Wheeler \cite{PhysRev.108.1063}, and Zerilli \cite{PhysRevLett.24.737} to decompose
the metric perturbations according to their transformation properties
under two-dimensional rotations.
Although Regge, Wheeler and Zerilli considered
the perturbations of the Schwarzschild space-time in GR, the formalism
 depends on the properties of spherical symmetry and then can
be applied to $f(R)$ gravity as well.

Let us denote the metric slightly perturbed from a static spherically symmetric
 spacetime by $g_{\mu\nu}=g_{\mu\nu}^{0}+h_{\mu\nu}$,  where $h_{\mu\nu}$ represents
infinitesimal quantities. In the lowest, linear approximation, the perturbations are supposed to be very smaller with respect to
 the background, that is  $g_{\mu\nu}^{0}>>h_{\mu\nu}$. Then, under two-dimensional
rotations on a sphere, $h_{tt},h_{tr}$ and $h_{rr}$ transform as scalars, $h_{ta}$
and $h_{ra}$ transform as vectors and $h_{ab}$ transforms as a tensor
($a,b$ are either $\theta$ or $\phi$). Any scalar  quantity $\Phi$  can be expressed in terms of the spherical harmonics $Y_{\ell m}(\theta,\phi)$
 \begin{equation}
\Phi(t,r,\theta,\phi)=\sum_{\ell,m}\Phi_{\ell m}(t,r)Y_{\ell m}(\theta,\varphi).\label{scalar-decomposition}
\end{equation}
In the   spherically symmetric spacetimes the solution will be independent of the index  $m$,
therefore this subscript can be omitted and we take into account only the index $\ell$ which represents  the multipole  number,
which arises from the separation of angular variables by the
expansion into spherical harmonics
\begin{equation} \Delta_{\theta, \phi}Y_{\ell}(\theta, \phi)=-\ell(\ell+1)Y_{\ell}(\theta, \phi),
\end{equation}
exactly in the same way as it happens for the hydrogen
atom problem in quantum mechanics when dealing with
the Schr\"odinger equation. Any vector $V_{a}$ can be decomposed into a divergence part
and a divergence-free part as follows:
 \begin{equation}
V_{a}(t,r,\theta,\phi)=\nabla_{a}\Phi_{1}+E_{a}^b\nabla_b\Phi_{2},
\end{equation}
where $\Phi_{1}$ and $\Phi_{2}$ are two scalars and $E_{ab}\equiv\sqrt{\det\gamma}~\epsilon_{ab}$
with $\gamma_{ab}$ being the two-dimensional metric on the sphere
and $\epsilon_{ab}$ being the totally anti-symmetric symbol with
$\epsilon_{\theta \varphi}=1$. Here $\nabla_{a}$ represents the covariant derivative with respect to the metric $\gamma_{ab}$.
Since $V_{a}$ is a two-component vector, it
is completely specified by the quantities $\Phi_{1}$ and $\Phi_{2}$. Then we can
apply the scalar decomposition (\ref{scalar-decomposition}) to $\Phi_{1}$
and $\Phi_{2}$ to decompose the vector quantity $V_a$ into spherical
harmonics.

Finally, any symmetric tensor $T_{ab}$ can be decompose as
 \begin{equation}
T_{ab}(t,r,\theta,\phi)=\nabla_{a}\nabla_{b}\Psi_{1}+\gamma_{ab}\Psi_{2}+\frac{1}{2}\left(E_{a}{}^{c}\nabla_{c}\nabla_{b}\Psi_{3}+
E_{b}{}^{c}\nabla_{c}\nabla_{a}\Psi_{3}\right),
\end{equation}
where $\Psi_{1},~\Psi_{2}$ and $\Psi_{3}$ are scalars. Since $T_{ab}$ has three
independent components, $\Psi_{1},~\Psi_{2}$ and $\Psi_{3}$ completely
specify $T_{ab}$. Then we can again apply the scalar decomposition
(\ref{scalar-decomposition}) to $\Psi_{1},~\Psi_{2}$ and $\Psi_{3}$
to decompose the tensor quantity $T_{ab}$ into spherical harmonics. We
refer to the variables accompanied by $E_{ab}$ by
 odd-type variables and the others by even-type
variables. What makes these decompositions useful is that, in the linearized equations
of motion (or equivalently, in the second order action) for $h_{\mu\nu}$,
odd-type  and even-type perturbations are completely decoupled. This fact  reflects the invariance of the background spacetime
under parity transformations.
Therefore, one can study odd-type perturbations and even-type ones
separately as we will do in the following.

\subsection{Perturbations in $f(R)$ gravity}

\centerline{The odd modes}

It is well known that there are two classes of vector
spherical harmonics (polar and axial) which are build out of combinations of
the Levi-Civita volume form and the gradient operator acting on the scalar
spherical harmonics. The difference between the two families is their parity.
Under the parity operator $\pi$ a spherical harmonic with index $\ell$ transforms as
$(-1)^\ell$, the polar class of perturbations transform under parity in the same way,
as $(-1)^\ell$ and the axial perturbations as $(-1)^{\ell+1}$.

Using the Regge-Wheeler formalism, the odd-type metric perturbations
can be written as
 \begin{eqnarray}
 &  & h_{tt}=0,~~~h_{tr}=0,~~~h_{rr}=0,\\
 &  & h_{ta}=\sum_{\ell, m}h_{0,\ell m}(t,r)E_{ab}\partial^{b}Y_{\ell m}(\theta,\varphi),\\
 &  & h_{ra}=\sum_{\ell, m}h_{1,\ell m}(t,r)E_{ab}\partial^{b}Y_{\ell m}(\theta,\varphi),\\
 &  & h_{ab}=\frac{1}{2}\sum_{\ell, m}h_{2,\ell m}(t,r)\left[E_{a}^{~c}\nabla_{c}\nabla_{b}Y_{\ell m}(\theta,\varphi)+E_{b}^{~c}\nabla_{c}\nabla_{a}Y_{\ell m}(\theta,\varphi)\right].
\end{eqnarray}
Using the gauge transformation $x^{\mu}\to x^{\mu}+\xi^{\mu}$, where $\xi^{\mu}$ are infinitesimal, we can show that not all the metric
perturbations are physical and some of them can be set
to vanish. For the odd-type perturbation,
we can consider the following gauge transformation:
 \begin{equation}
\xi_{t}=\xi_{r}=0,~~~\xi_{a}=\sum_{\ell m}\Lambda_{\ell m}(t,r)E_{a}^{~b}\nabla_{b}Y_{\ell m},
\end{equation}
where $\Lambda_{\ell m}$  can always set $h_{2,\ell m}$ to vanish
(Regge-Wheeler gauge). By this procedure, $\Lambda_{\ell m}$ is
completely fixed and there is no remaining gauge degrees of freedom. Then, after substituting the metric
into the action (\ref{action2}) and performing integrations by parts,
we find that the action for the odd modes becomes
\begin{equation}
S_{{ odd}}=\frac{1}{2\kappa} \sum_{\ell,m}\int dt\, dr\,{\cal L}_{{ odd}}=\frac{1}{4\kappa} \sum_{\ell,m}\int dt\, dr\,j^{2}\bigg[\phi{\left({\dot{h}_{1}}-h_{0}'\right)}^{2}+\frac{4h_{0}{\dot{h}_{1}} \phi}{r}+ \frac{h_{0}^{2}}{r^{2}}\left[2r\phi'+2\phi +\frac{(j^2-2)\phi}{B} \right]-\frac{\,(j^{2}-2)\,B\, \phi\, h_{1}^{2}}{r^{2}}\bigg],\label{odd-action}
\end{equation}
where we neglect the suffix $\ell$ for the fields, and $j^{2}=\ell\,(\ell+1)$. Variation of (\ref{odd-action}) with respect to $h_{0}$ yields
 \begin{equation}
[\phi(h_{0}'-\dot{h}_{1})]'= \frac{1}{r^{2}}\left[r\phi'+j^{2}\phi+ \frac{(j^2-2)\phi}{2B}\right]\, h_{0}+\frac{2 \phi\,{\dot{h}_{1}}}{r}\,,\label{cons}
\end{equation}
which cannot be solved for $h_{0}$. Let us now rewrite the above action as
\begin{equation}
{L}_{{ odd}}=\frac{ j^{2}\, \phi}{2}{\left({\dot{h}_{1}}-h_{0}'+\frac{2\,{h_0}}{r}\right)}^{2}-\frac{j^{2}(\phi+r\phi')\,h_{0}{}^2}{r^2}+\frac{2j^{2}\,h_{0}^{2}}{r^{2}}\left[r\phi'+\phi +\frac{(j^2-2)\phi}{2B} \right]-\frac{j^{2}\,(j^{2}-2)\,B\, \phi\, h_{1}^{2}}{2r^{2}}.\label{eq:Lodd2}
\end{equation}
so that all the terms containing $\dot{h}_{1}$ are inside the first squared term. Using a Lagrange multiplier $Q$, we can rewrite Eq.~(\ref{eq:Lodd2})
as follows
\begin{equation}
{L}_{{  odd}}=\frac{j^{2}\, \phi}{2}\left[2\, Q\left(\dot{h}_{{1}}-h'_{{0}}+{\frac{2\,h_{{0}}}{r}}\right)-Q^{2}\right]-\frac{j^{2}(\phi+r\phi')h_{0}{}^{2}}{r^2}+\frac{2j^{2}\,h_{0}^{2}}{r^{2}}\left[r\phi'+\phi +\frac{(j^2-2)\phi}{2B} \right]-\frac{j^{2}\,(j^{2}-2)\,B\, \phi\, h_{1}^{2}}{2r^{2}}\,.\label{eq:Lodd3}
\end{equation}
Eq. (\ref{eq:Lodd3}) shows that both fields $h_{0}$ and $h_{1}$ can be integrated out by using
their own equations of motion, which can be written as
\begin{eqnarray}
h_{1} & = & -\frac{r^2\,\dot{Q}}{(j^{2}-2)B}\,,\label{eq:oddh1}\\
h_{0} & = & \frac{r}{\phi(j^{2}-2)}\,[(\phi+r\phi')\, Q+r\, \phi\, Q']\:.\label{eq:oddh0}
\end{eqnarray}
 These relations link the physical modes $h_{0}$ and $h_{1}$ to
the auxiliary field $Q$. Once $Q$ is known also $h_{0}$ and $h_{1}$
are. After substituting these expressions into the Lagrangian and performing
an integration by parts for the term proportional to $Q'\, Q$, one finds the Lagrangian
in the canonical form
\begin{equation}
{L}_{{odd}}=\frac{j^2r^2\phi}{2(j^2-2)B}\,\dot{Q}^{2}-\frac{j^2B\, \phi\,r^{2}}{2(j^2-2)}\,Q'^{2}-\mu_1{}^{2}\, Q^{2}\,,\label{eq:LoddF}
\end{equation}
 where
\begin{eqnarray}
\mu_1{}^{2}&=&\frac{j^2\Big[j^2\phi^2-Br^2\phi\phi''+2B\phi^2-r^2\phi\phi'B'+r^2B\phi'^2-2\phi^2-2r\phi^2 B'\Big]}{2\phi (j^2-2)}\,.
\end{eqnarray}
From Eq.~(\ref{eq:LoddF}), we can derive the no ghost conditions
\[
j^2\geq2\,,\qquad{  \mbox{and} \, \,\qquad}B\geq0\,.
\]
For solutions proportional to $e^{i(\omega t-kr)}$ with large
$k$ and $\omega$, we have the radial dispersion relation
\[
\omega^{2}=B^2\, k^{2}\,,
\]
 where we made use of the background equations of motion. Finally
the expression for the radial speed reads

\[
c_{{ odd}}^{2}=\left(\frac{dr_{*}}{d\tau}\right)^2=1\,,
\]
 where we used the radial tortoise coordinate ($dr_{*}^{2}=dr^{2}/B$)
and the proper time ($d\tau^{2}=B\, dt^{2}$).

%%%%%%%%%%%%%%%%%%%%%%%%%%%% Section 7 %%%%%%%%%%%%%%%%%%%%%%%%%%%%%
\subsection{Black hole stability: Geodesic}\label{S66}
%%%%%%%%%%%%%%%%%%%%%%%%%%%%%%%%%%%%%%%%%%%%%%%%%%%%%%%%%%%%%%%%%%%%
 The trajectories of a test particle in a gravitational field are described   by the geodesic equations
 \begin{equation}\label{ge}
 {d^2 x^\sigma \over d\lambda^2}+ \left\{^\sigma_{ \mu \nu} \right \}
 {d x^\mu \over d\lambda}{d x^\nu \over d\lambda}=0,
 \end{equation}
 where $\lambda$ is an affine  parameter along the geodesic. The
  geodesic  deviation takes the form \cite{1992ier..book.....D}
  \begin{equation} \label{ged}
 {d^2 \xi^\sigma \over d\lambda^2}+ 2\left\{^\sigma_{ \mu \nu} \right \}
 {d x^\mu \over d\lambda}{d \xi^\nu \over ds}+
 \left\{^\sigma_{ \mu \nu} \right \}_{,\ \rho}
 {d x^\mu \over d\lambda}{d x^\nu \over d\lambda}\xi^\rho=0,
 \end{equation}
with $\xi^\rho$ being the deviation 4-vector. Applying (\ref{ge}) and (\ref{ged})  into (\ref{met}) we get for the geodesic equations \begin{equation}
{d^2 t \over d\lambda^2}=0, \qquad {1 \over 2} B'(r)\left({d t \over
d\lambda}\right)^2-r\left({d \phi \over d\lambda}\right)^2=0, \qquad {d^2
\theta \over d\lambda^2}=0,\qquad {d^2 \phi \over d\lambda^2}=0,\end{equation} and for
the geodesic deviation \begin{eqnarray}\label{ged1} && {d^2 \xi^1 \over d\lambda^2}+B(r)B'(r) {dt \over d\lambda}{d
\xi^0 \over d\lambda}-2r B(r) {d \phi \over d\lambda}{d \xi^3 \over
d\lambda}+\left[{1 \over 2}\left(B'^2(r)+B(r) B''(r)
\right)\left({dt \over d\lambda}\right)^2-\left(B(r)+rB'(r)
\right) \left({d\phi \over d\lambda}\right)^2 \right]\xi^1=0, \nonumber\\
&&  {d^2 \xi^0 \over
d\lambda^2}+{B'(r) \over B(r)}{dt \over d\lambda}{d \zeta^1 \over d\lambda}=0,\qquad {d^2 \xi^2 \over d\lambda^2}+\left({d\phi \over d\lambda}\right)^2
\xi^2=0, \qquad \qquad  {d^2 \xi^3 \over d\lambda^2}+{2 \over r}{d\phi \over d\lambda} {d
\xi^1 \over d\lambda}=0, \end{eqnarray} where $B(r)$ is defined by the metric (\ref{me}) or (\ref{me1}),
$B'(r)=\displaystyle{dB(r) \over dr}$. Using
the circular orbit  \begin{equation} \theta={\pi \over 2}, \qquad
{d\theta \over d\lambda}=0, \qquad {d r \over d\lambda}=0,
\end{equation}
we get
\begin{equation}
 \left({d\phi \over d\lambda}\right)^2={B'(r)
\over r(2B(r)-rB'(r))}, \qquad \left({dt \over
d\lambda}\right)^2={2 \over 2B(r)-rB'(r)}. \end{equation}

Eqs. (\ref{ged1}) can be rewritten as
\begin{eqnarray} \label{ged2} &&  {d^2 \xi^1 \over d\phi^2}+B(r)B'(r) {dt \over
d\phi}{d \xi^0 \over d\phi}-2r B(r) {d \xi^3 \over
d\phi} +\left[{1 \over 2}\left(\eta'^2(r)+\eta(r) \eta''(r)
\right)\left({dt \over d\phi}\right)^2-\left(\eta(r)+r\eta'(r)
\right)  \right]\zeta^1=0, \nonumber\\
&&{d^2 \xi^2 \over d\phi^2}+\xi^2=0, \qquad {d^2 \xi^0 \over d\phi^2}+{B'(r) \over
B(r)}{dt \over d\phi}{d \xi^1 \over d\phi}=0,\qquad {d^2 \xi^3 \over d\phi^2}+{2 \over r} {d \xi^1 \over
d\phi}=0. \end{eqnarray}
The second equation  of (\ref{ged2}) shows that it is a simple harmonic motion which  means that the motion in
the plan $\theta=\pi/2$ is stable. Now  the solutions of the remaining equations of (\ref{ged2}) are given by  \begin{equation} \label{ged3}
\xi^0 = \zeta_1 e^{i \sigma \phi}, \qquad \xi^1= \zeta_2e^{i \sigma
\phi}, \qquad and \qquad \xi^3 = \zeta_3 e^{i \sigma \phi},\end{equation}  where
$\zeta_1, \zeta_2$ and $\zeta_3$ are constants, and  the variable $\phi$ has  to be determined. Substituting (\ref{ged3}) in
(\ref{ged2}),  we get  \begin{equation} \label{con1}  \displaystyle\frac{3BB'-\omega^2B'-2rB'^2+rBB''}{B'}>0, \end{equation} which is the stability condition for any charged static spherically symmetric spacetime. The condition (\ref{con1}) for the black holes (\ref{me}) and (\ref{me1}) can be rewritten as  \begin{equation}
r+\displaystyle{2q^2 \over 5M}>0, \qquad \qquad
r+12M>0, \,\qquad \mbox{and}  \qquad 1+\displaystyle{4r^3\Lambda  \over 31M}>0,\end{equation}
which are the  stability conditions according to the values of the parameters $\Lambda$, $M$ and $q$.

\section{ Discussion and conclusions }\label{S77}

Spherically symmetric  spacetimes constitute an essential part of black hole physics because all the fundamental properties of the black holes can be explained and can further be used to recognize and hence generalize  in any eligible  more general scenario \cite{Chakraborty:2016lxo}. In this paper, we discussed two main issues. In the first part, we focused on a spherically symmetric spacetime in the framework of  $f(R)$ gravitational theories. We  derived new black hole charged solutions for the specific forms  $f(R)=R-2\alpha\sqrt{R}$ and $f(R)=R-2\alpha\sqrt{R-8\Lambda}$. The main merits of these black holes are the fact that they depend on the dimensional parameter $\alpha$ and have dynamical Ricci scalar, i.e., $R=\frac{1}{r^2}$ for the first model of $f(R)$ and $R=\frac{8r^2\Lambda +1}{r^2}$ for the second one. These solutions are new and cannot reduce to the standard solutions of GR due to the fact that the parameter $\alpha$ is not allowed to have a zero value. We calculate the scalar invariant of those black holes and found that the Kretschmann and Ricci tensor square invariants depending on the dimensional parameter $\alpha$. All of the  invariants show true singularity at $r=0$.

 In the second part, we study the thermodynamical properties of these black holes to extract more physical information from them. The first important thing in $f(R)$ gravity is the fact that entropy is not always proportional to the area of the horizon \cite{Cvetic:2001bk,Nojiri:2001fa,PhysRevD.65.023521}. { We have shown that, for some constraint  on the parameter $0<\alpha<0.5$, we have a positive value of the entropy.} However, for the black hole solution (\ref{sol1}) there is a region  in which  the entropy has a negative value \cite{Cvetic:2001bk,Nojiri:2001fa,PhysRevD.65.023521,Clunan:2004tb}. This is not the first time that  a black hole with negative entropy is found. Several  black holes with negative   entropy have been found as well as in charged Gauss-Bonnet (A)dS gravity \cite{Cvetic:2001bk,Nojiri:2001fa,PhysRevD.65.023521}. As our calculations show,  negative entropy may be interpreted as a region where the parameter $\alpha$ has   transitions  into  forbidden regions related to some   phase transition.  The complete understanding of gravitational entropy of non-trivial solution in the framework of $f(R)$ gravitational theories remains the subject of future research.

 We also  calculated the thermodynamical quasi-local  energy and showed  that it has a positive value when $0<\alpha<0.5$.   Moreover, we calculated the Hawking temperature and have shown that it also depends on the parameter $\alpha$. Also, we have shown that the Hawking temperature has always positive   value when   $0.5>\alpha>0$ for the  black hole  (\ref{sol}) and (\ref{sol1}).   In fact, this is the case presented in Fig. \ref{Fig:3}\subref{fig:3a} for the $\alpha>0$ regime. As for the black hole (\ref{sol1}) the Hawking temprature always show a positive value as Fig. \ref{Fig:3}\subref{fig:3b}. Finally, We o have calculated the Gibb's free energy and show that our black hole (\ref{sol}) is globally stable when $0<\alpha<0.5$. However, the black hole (\ref{sol1}) is not globally stable when $r<2.73$ and become stable when $r>2.73$. The main reason that makes  this black  is not stable comes from the contribution of entropy which has a negative value in the region $r<2.73$ as Fig. \ref{Fig:2}\subref{fig:2b} shows.  The results obtained here, together with other results in the literature, seem to indicate that the thermodynamical origin of $f(R)$ gravitational theories, when horizons are present, has a broad of validation. To confirm this statement we need to know more about the novel black holes derived in this paper. This will be done in  future studies.

Finally, we have studied the linear perturbations around the static spherically symmetric charged spacetime derived in $f(R)$ gravity. Due to the fact that $f(R)$ is a fourth order theory,  we have rewritten its Lagrangian as a Ricci scalar coupled with a scalar field to make the study of perturbation more practice. We have  derived the
gradient instability condition for our black holes using the odd-type modes. Furthermore, we   have calculated the radial propagation speed and showed  that it is equal one. To make the picture more complete,  we have derived  the stability conditions using also the geodesic deviation for the black holes. These conditions are different with respect to  the charged black hole of GR, the Reissner-Nordstr\"om spacetime. This difference is due to the fact that the charged black hole derived in this study is a solution in the context of $f(R)$ only and cannot be reduced to GR.

\section*{Acknowledgments}
GN would like to thank S.  Odintsov for useful discussion. SC is supported in part by the INFN sezione di Napoli, {\it iniziative specifiche QGSKY and MOONLIGHT2}. The article is also based upon work from COST action CA15117 (CANTATA), supported by COST (European Cooperation in Science and Technology).
%%%%%%%%%%%%%%%%%%%%%%%%%%%%%%%%%%%%%%%%%%%%%%%%%%%%%%%%%%%%%%%%%%%%%%%%%%%%%%%%%%%%%%
%\bibliographystyle{apsrev}
%\bibliography{Ref}

\end{document}